\documentclass[twocolumn,showpacs,preprintnumbers,amsmath,amssymb,superscriptaddress,prb]{revtex4}

\usepackage{graphicx}
\usepackage{dcolumn}
\usepackage{bm}
\usepackage{braket}

\begin{document}

\preprint{APS/123-QED}

\title{$^{51}$V-NMR study on the $S$=1/2 square lattice antiferromagnet K$_2$V$_3$O$_8$}

\author{H. Takeda}
\affiliation{Institute for Solid State Physics, University of Tokyo, 5-1-5 Kashiwanoha, Kashiwa, Chiba 277-8581, Japan}%
\author{H. Yasuoka}%
\affiliation{Los Alamos National Laboratory, Los Alamos, New Mexico 87545, USA}%

\author{M. Yoshida}%
\affiliation{Institute for Solid State Physics, University of Tokyo, 5-1-5 Kashiwanoha, Kashiwa, Chiba 277-8581, Japan}%

\author{M. Takigawa}%
\affiliation{Institute for Solid State Physics, University of Tokyo, 5-1-5 Kashiwanoha, Kashiwa, Chiba 277-8581, Japan}%

\author{N. J. Ghimire}%
\affiliation{Los Alamos National Laboratory, Los Alamos, New Mexico 87545, USA}%
\affiliation{Department of Physics and Astronomy, The University of Tennessee, Knoxville, Tennessee 37996, USA}%
\affiliation{Materials Science and Technology Division, Oak Ridge National Laboratory, Oak Ridge, Tennessee 37831, USA}%

\author{D. Mandrus}%
\affiliation{Materials Science and Technology Division, Oak Ridge National Laboratory, Oak Ridge, Tennessee 37831, USA}%
\affiliation{Department of Materials Science and Engineering, The University of Tennessee, Knoxville, Tennessee 37996, USA}%

\author{B. C. Sales}%
\affiliation{Materials Science and Technology Division, Oak Ridge National Laboratory, Oak Ridge, Tennessee 37831, USA}%

\date{\today}

\begin{abstract}
Static and dynamic properties of the quasi-two-dimensional antiferromagnet K$_2$V$_3$O$_8$ have been investigated by $^{51}$V-NMR experiments on nonmagnetic V$^{5+}$ sites. Above the structural transition temperature $T_{\rm{S}}$ = 115 K, NMR spectra are fully compatible with the $P4bm$ space group symmetry. The formation of superstructure below $T_{\rm{S}}$ causes splitting of the NMR lines, which get broadened at lower temperatures so that individual peaks are not well resolved. Evolution of NMR spectra with magnetic field along $c$-axis below the magnetic transition temperature $T_{\rm{N}} \sim 4$ K is qualitatively consistent with a simple N\'{e}el order and a spin flop transition. However, broad feature of the spectra does not rule out possible incommensurate spin structure. The spin-lattice relaxation rate $1/T_1$ below $T_{\rm{N}}$ shows huge enhancement for a certain range of magnetic field, which is independent of temperature and attributed to cross relaxation due to anomalously large nuclear spin-spin coupling between V$^{5+}$ and magnetic V$^{4+}$ sites. The results indicate strong gapless spin fluctuations, which could arise from incommesurate orders or complex spin textures.        
\end{abstract}

\maketitle

\section{introduction}
Spin systems with non-centrosymmetric crystal structure often exhibit novel magnetic phenomena induced by spin-orbit coupling.\cite{Dzyaloshinski1, Rosler, Rosler2, Bogdanov}  For example, the Dzyaloshinskii-Moriya (DM) interaction \cite{Dzialoshinski, Moriya0} generated by spin-orbit coupling plays an important role to stabilize non-collinear spin structures such as helical or canted antiferromagnetic order.  These non-collinear spin structures produce higher order spin degrees of freedom such as scalar or vector spin chiralities, whose coupling to external magnetic field or crystal lattice results in peculiar transport or cross correlation properties such as non-trivial magnetoresistance, anomalous Hall effect, and multiferroics. \cite{Nagaosa1, Tokura1}
In particular, cubic non-centrosymmetric crystals such as MnSi or FeGe have attracted great attentions due to their exotic magnetic structure under magnetic field.\cite{Muhlbauer, Uchida, Nagaosa}  The absence of inversion symmetry in these systems allows chiral magnetic order with a long period due to competition between ferromagnetic and DM interactions in zero field.  Application of modest magnetic fields induces a transition into an exotic magnetic structure, where a lattice of topological spin texture called skyrmions appears before reaching the saturated ferromagnetic state at a higher magnetic field. 

\begin{figure}[t]
\includegraphics[width=8.5cm]{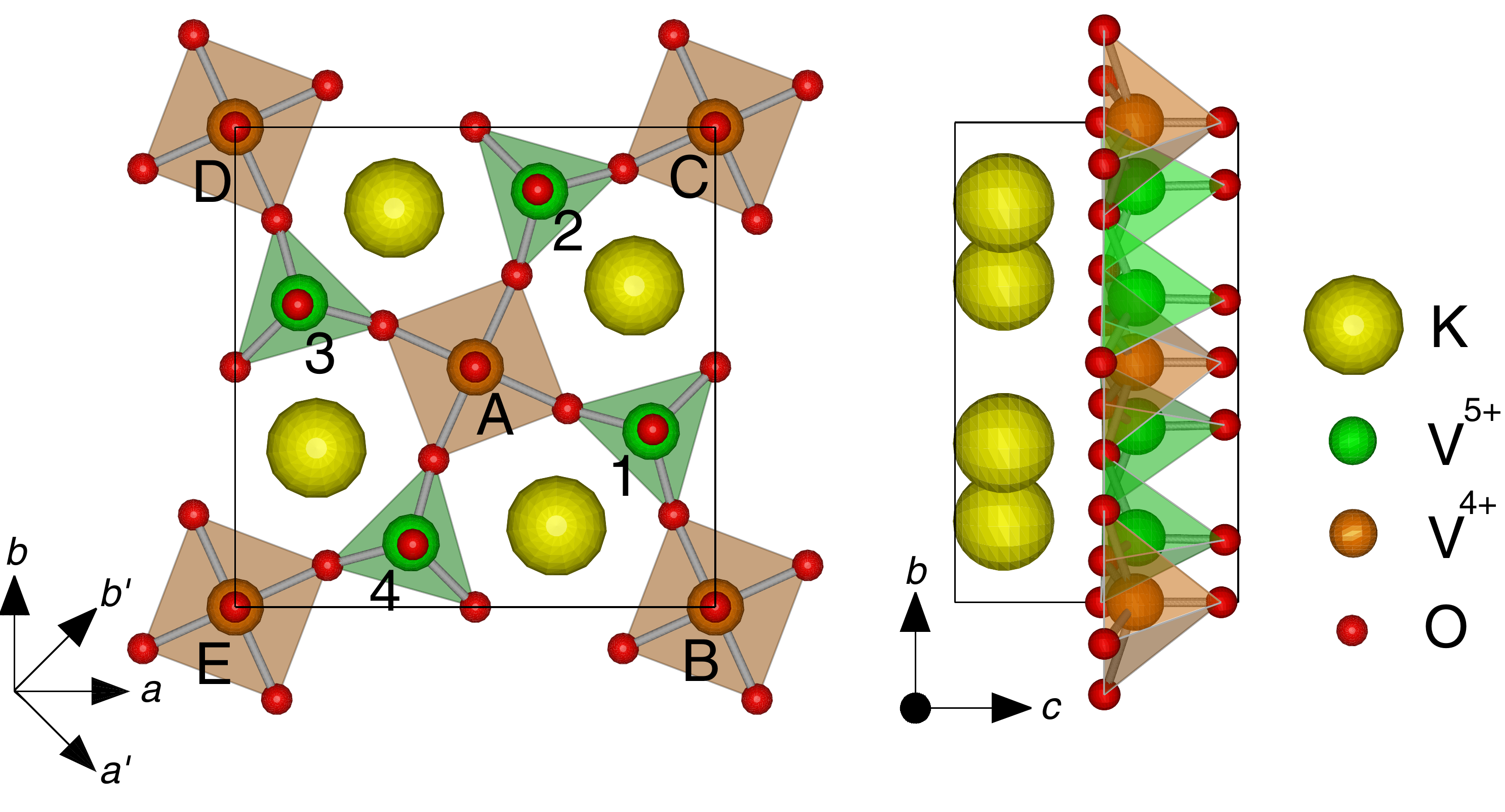}
\caption{\label{fig:crystal} (Color online) Crystal structure of K$_2$V$_3$O$_8$ drawn by VESTA\cite{Momma} in the high temperature phase with the space group $P4bm$ viewed along the $c$-axis (left panel) and along the $a$-axis (right panel).  A unit cell is shown by the solid lines.  $a'$ and $b'$ are the directions rotated by 45$^{\circ}$ from the $a$ and $b$ axes in the $ab$-plane.} 
\end{figure}

Similar spin texture or chiral magnetic phases were also predicted theoretically for certain antiferromagnets with a non-centrosymmetric structure.  Bogdanov {\it et al}. predicted a modulated magnetic structure in the spin 1/2 antiferromagnet K$_2$V$_3$O$_8$ with a quasi-two dimensional square-lattice.\cite{Bogdanov}  K$_2$V$_3$O$_8$ crystalizes in the tetragonal Fresonoite-type structure with the space group $P4bm$ as shown in Fig. \ref{fig:crystal}, which consists of alternating V-O and K layers. A V-O layer is formed by two types of VO polyhedra, namely VO$_5$ pyramids with V$^{4+}$ are connected to VO$_4$ tetrahedra containing V$^{5+}$ by corner sharing.\cite{Lumsden1, Liu}  K$_2$V$_3$O$_8$ undergoes an antiferromagnetic transition at $T_{\rm{N}} \sim$4 K.  Neutron diffraction measurements revealed a simple two-sublattice N\'{e}el order with spins aligned along the $c$-axis at zero magnetic field.\cite{Lumsden1}  Application of external magnetic field induces intriguing spin reorientations.\cite{Lumsden1}  While a spin flop transition occurs at the field of 0.85~T along the $c$-axis, fields along the $a$-axis greater than 0.65~T cause the spins to rotates continuously from the $c$-axis onto the $ab$-plane while remaining perpendicular to the field. Such spin reorientation behavior was explained by combination of the $c$-component of the anti-symmetric DM interaction $D_{c}({\bf {S}}_{1} \times {\bf {S}}_{2})_{c}$ and the symmetric easy-axis anisotropic interaction.\cite{Lumsden1} Substantial enhancement of the thermal conductivity has been reported for the fields above these critical values, which is attributed to the gapless spin waves in the new ground state in magnetic fields.\cite{Sales}   

Based on a phenomenological continuum model, Bogdanov {\it et al}. then pointed out that the in-plane component of the DM interaction can stabilize a modulated structure with topological defects such as vortices or skyrmions in the vicinity of the spin flop transition.\cite{Bogdanov}  However, such a structure has not been reported so far.

Further experiments revealed a structural transition at  $T_{\rm{S}} \sim$115 K.\cite{JChoi, Rai, KChoi, Chakoumakos}  While the splitting of of some phonon modes observed by infrared spectroscopy provides evidence for local distortion of VO$_5$ pyramids,\cite{JChoi, Rai, KChoi} the magnetic susceptibility shows no anomaly at $T_{\rm{S}}$.\cite{Lumsden1, Liu}  X-ray diffraction measurements detected weak superlattice reflections indicating a $3 \times 3 \times 2$ supercell, whereas the precise structure has not been determined.\cite{Chakoumakos}  Such a lattice modulation could result in an incommensurate spin structure due to corresponding modulation of exchange interactions,\cite{Zaliznyak}  although incommensurate Bragg peaks have not been detected by neutron scattering measurements.\cite{Lumsden1, Lumsden2}  On the other hand, large magneto-optical effects on the V$^{4+}$$d \rightarrow d$ onsite excitation spectrum indicates field-induced local distortion and strong spin-lattice coupling.\cite{Rai}  Raman spectroscopy also detected mixing between the spin wave excitations and phonon vibrations.\cite{KChoi}

In this paper, we discuss microscopic structural and magnetic properties of K$_2$V$_3$O$_8$ based on $^{51}$V NMR experiments on V$^{5+}$ sites. We observed splitting of the NMR lines below $T_{\rm{S}}$, which can be ascribed to the formation of superlattice and associated local distortion of the V$^{4+}$O$_5$ pyramids. The lines get further broadened with decreasing temperature and severely overlap with each other, resulting in a broad NMR spectra even above $T_{\rm{N}}$. Evolution of the spectral shape below $T_{\rm{N}}$ in magnetic fields along $c$-axis can be explained qualitatively by the spin flop transition with a simple N\'{e}el order as proposed by previous studies. However, broad feature of the NMR spectra does not rule out possibility for incommensurate spin structures. The most remarkable observation in our work is the huge enhancement of the spin-lattice relaxation rate $1/T_1$ in a certain range of magnetic field, which is independent of temperature below $T_{\rm{N}}$. This is explained by the V$^{5+}$- V$^{4+}$ cross-relaxation process with unusually large nuclear spin-spin coupling, indicating dense gapless spin fluctuations in high magnetic fields. Such spin fluctuations could arise in magnetically ordered states with incommensurate structures or complex spin textures.   
 
\section{Experimental Procedure}
A single crystal of K$_2$V$_3$O$_8$ with the size 1.7$\times$3.0$\times$0.4 mm$^3$ used in this work was prepared as described in Ref. \onlinecite{Lumsden1}.  $^{51}$V NMR measurements were performed using a pulse NMR spectrometer with a double axis goniometer for precise alignment of the crystal in a magnetic field. NMR spectra were obtained by summing the Fourier transform of the spin echo signal recorded at equally spaced frequencies in a fixed magnetic field. The spin lattice relaxation rate $1/T_1$ was measured by the saturation or inversion recovery methods with the excitation-pulse-width less than 1~$\mu$s. The bandwidth of the excitation pulse was broad enough to saturate all the quadrupole split lines, therefore, the recovery of nuclear magnetization after the excitation pulse always followed an exponential function with a single value of $T_1$.  

The NMR resonance frequencies are generally given by the following nuclear spin Hamiltonian 
\begin{eqnarray}{\label{nu_hamiltonian}}
{\mathcal{H}}_{\rm I} & = & h \gamma {\bf{I}}\cdot{\bf{H}}_{\rm{loc}}+ \sum_{\alpha, \beta} V_{\alpha \beta} Q_{\alpha \beta}  \\
V_{\alpha \beta} & = & \frac{\partial^2 V}{\partial x_{\alpha}\partial x_{\beta}}, \nonumber \\
Q_{\alpha \beta} & = & \frac{eQ}{6I(2I-1)} \left\{ \frac{3}{2} \left( I_{\alpha}I_{\beta} +  I_{\beta}I_{\alpha}\right)-\delta_{\alpha \beta}I(I+1) \right\},  \nonumber
\end{eqnarray}
where {\bf I} is the nuclear spin and $I = 7/2$ for $^{51}$V nuclei. The first term represents the Zeeman interaction between the nuclear magnetic moment $h \gamma {\bf{I}}$ and the local magnetic field ${\bf{H}}_{\rm{loc}}$ acting on a nucleus, where $\gamma = 11.1988$MHz/T is the gyromagnetic ratio of $^{51}$V nuclei. This term splits the energy level into $2I+1$ eigenstates of $|I_{z} = m \rangle$, where $z$ is the direction of ${\bf{H}}_{\rm{loc}}$, generating a single NMR line at the frequency $\nu = \gamma H_{\rm{loc}}$. 

The resonance line is then split by the second term, which represents the quadrupole interaction between the nuclear quadrupole moment tensor ${\bf {Q}}$ and the electric field gradient (EFG) tensor ${\bf {V}}$ defined with respect to an appropriate crystalline coordinate frame. 
Since the Zeeman energy is much larger than the quadrupole interaction in our experiments as shown below, it is sufficient to consider the latter up to the first order in perturbation. Then the frequencies of the quadrupole split line for the transition $I_{z}=m\leftrightarrow m-1$ ($m=-I+1,\cdots,I$) is given by
\begin{eqnarray}
\nu_{m \leftrightarrow m-1} & = & \gamma H_{\rm{loc}} + \frac{3V_{zz}eQ}{h2I(2I-1)} (m-\frac{1}{2}) \\
V_{zz} & = & {\bf{h}}\cdot{\bf {V}}\cdot{\bf{h}} ,
\end{eqnarray}
where ${\bf{h}}$ is the unit vector along ${\bf{H}}_{\rm{loc}}$ and $V_{zz}$ is the EFG along the direction of ${\bf{H}}_{\rm{loc}}$. Thus a single vanadium site produces seven equally spaced NMR lines. The value of  $H_{\rm{loc}}$ can be obtained from the frequency of the central line
\begin{equation}{\label{centerline}}
\gamma H_{\rm{loc}} = \nu_{1/2 \leftrightarrow -1/2} , 
\end{equation}
while the EFG tensor is determined from the spacing between a pair of satellite lines,
\begin{equation}{\label{Qcoupling}}
\nu^{\rm{Q}}_{zz} = {\bf{h}}\cdot{\bm{\nu}}^{\rm{Q}}\cdot{\bf{h}} = \frac{\nu_{m \leftrightarrow m-1}-\nu_{-m+1 \leftrightarrow -m}}{2m-1} ,
\end{equation}
where the quadrupole coupling tensor is defined as
\begin{equation} 
{\bm{\nu}}^{\rm{Q}} = \frac{3eQ{\bf{V}}}{h2I(2I-1)} .
\end{equation}   
 
The local field is composed of a macroscopic field, which is sum of the external field (${\bf{H}}_{\rm{ext}}$) and the Lorentz and demagnetization fields, and a microscopic hyperfine field (${\bf{H}}_{\rm{hf}}$) as
\begin{equation} 
{\bf{H}}_{\rm{loc}} = {\bf{H}}_{\rm{ext}} + \frac{4\pi}{3}\left({\bf{1}}-3{\bf{N}}\right)\cdot{\bf{M}}_{\rm{v}} + {\bf{H}}_{\rm{hf}} .
\end{equation} 
In the second term representing the sum of the Lorentz and demagnetization fields, ${\bf{M}}_{\rm{v}}$ is the magnetization per unit volume and ${\bf{N}}$ is the demagnetization tensor which can be determined from the shape of the crystal. After correcting for this term, the central line frequency in Eq.~(\ref{centerline}) then gives the value of $|{\bf{H}}_{\rm{ext}} + {\bf{H}}_{\rm{hf}}|$. The hyperfine field is produced by surrounding electron magnetic moments,
\begin{equation}{\label{hfcoupling}}
{\bf{H}}_{\rm{hf}}= \sum_i{\bf{A}}_i \cdot{\bm{\mu}}_i ,
\end{equation}
where ${\bf{A}}_i$ is the hyperfine coupling tensor of a nucleus to the magnetic moment ${\bm{\mu}}_i$ (with Bohr magneton $\mu_{\rm{B}}$ as a unit) at $i$ site.   

In the paramagnetic state, all the moments are uniform and induced by the external field, ${\bm{\mu}}_i = {\bf{M}}/N_{\rm{A}}\mu_{\rm{B}} = {\bm{\chi}}\cdot{\bf{H}}_{\rm{ext}} /N_{\rm{A}}\mu_{\rm{B}}$, where {\bf{M}} is the molar magnetization, $N_{\rm{A}}$ is the Avogadro's number, and ${\bm{\chi}}$ is the susceptibility tensor. Therefore, the hyperfine field is expressed as ${\bf{H}}_{\rm{hf}} = {\bf{A}}\cdot {\bm{\chi}}\cdot{\bf{H}}_{\rm{ext}} /N_{\rm{A}}\mu_{\rm{B}}$ with ${\bf{A}}=\sum_i{\bf{A}}_i$ and $\left| {\bf{H}}_{\rm{ext}} + {\bf{H}}_{\rm{hf}} \right| = \left| \left({\bf{1}}+{\bf{K}}\right)\cdot{\bf{H}}_{\rm{ext}} \right|$, where we defined the shift tensor 
\begin{equation}{\label{Kchi}}
{\bf{K}} = {\bf{A}} \cdot  {\bm{\chi}}/N_{\rm{A}}\mu_{\rm{B}} . 
\end{equation}
Since all components of ${\bf{K}}$ is of the order of  $10^{-2}$ or less as we will see below, it is sufficient to consider only the component of ${\bf{H}}_{\rm{hf}}$ parallel to ${\bf{H}}_{\rm{ext}}$. Therefore, the experimentally observed shift $K$ is given by 
\begin{equation}{\label{shift}}
K \equiv \frac{\left| {\bf{H}}_{\rm{ext}} + {\bf{H}}_{\rm{hf}} \right| - H_{\rm{ext}}}{H_{\rm{ext}}} \approx K_{zz} = {\bf{h}}\cdot{\bf{K}}\cdot{\bf{h}} .
\end{equation} 
The components of ${\bf{K}}$ can be determined by measuring $K$ for various field directions as discussed in section~\ref{para}. 

Certain components of ${\bf{K}}$ become zero by symmetry. Let us consider, for example, V$^{5+}$(1) site shown in Fig.~\ref{fig:crystal}. Since this site is on a mirror plane perpendicular to the $a'$ direction, it follows that if ${\bf{H}}_{\rm{ext}}$ is parallel (perpendicular) to $a'$, then ${\bf{H}}_{\rm{hf}}$ must be also parallel (perpendicular) to $a'$. Then the $a'b'$-, $a'c$-, $b'a'$-, and $ca'$-components of the shift tensor ${\bf{K}}_1$ at V$^{5+}$(1) sites should be zero. From now on, we will use the $a'b'c$ coordinate frame.
\begin{eqnarray}{\label{Ktensor}}
{{\bf{K}}}_1=\left(\begin{array}{ccc}
K_{a'a'} & 0 & 0\\
0 & K_{b'b'} & K_{b'c}\\
0 & K_{cb'} & K_{cc}
\end{array}\right).
\end{eqnarray}
Note that the same rule holds also for the quadrupole coupling tensor ${\bm{\nu}}^{\rm{Q}}$. Since other three V$^{5+}$ sites are generated by $C_4$ operation along the $c$-axis, the shift tensors at V$^{5+}$(2) $\sim$ V$^{5+}$(4) sites are obtained by successive application of $C_4$ to ${\bf{K}}_1$,   
\begin{eqnarray}
{{\bf{K}}}_2&= \left(\begin{array}{ccc}
K_{b'b'} & 0 & -K_{b'c}\\
0 & K_{a'a'} & 0\\
-K_{cb'} & 0 & K_{cc}
\end{array}\right), \\
{{\bf{K}}}_3&= \left(\begin{array}{ccc}
K_{a'a'} & 0 & 0\\
0 & K_{b'b'} & -K_{b'c}\\
0 & -K_{cb'} & K_{cc}
\end{array}\right), \\
{{\bf{K}}}_4&= \left(\begin{array}{ccc}
K_{b'b'} & 0 & K_{b'c}\\
0 & K_{a'a'} & 0\\
K_{cb'} & 0 & K_{cc}
\end{array}\right).
\end{eqnarray}

\section{Experimental Results and Analysis}
\subsection{Paramagnetic phase}{\label{para}}

\begin{figure}[t]
\includegraphics[width=8.5cm]{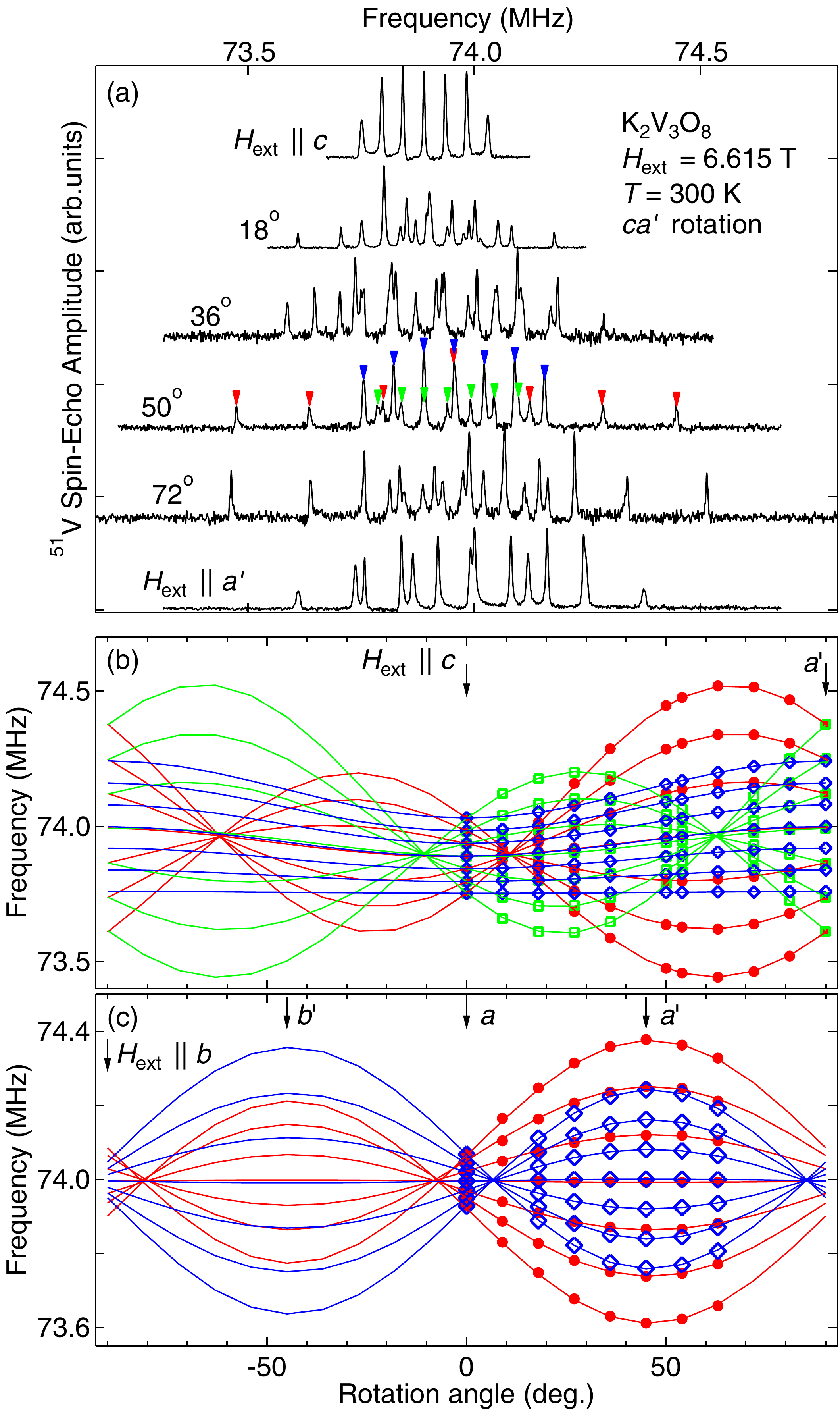}
\caption{\label{fig:300K_V_spc} (Color online) (a) $^{51}$V NMR spectra obtained at  300 K for the external field $H_{\rm{ext}}$ = 6.615 T applied in the $ca'$-plane. Three sets of spectra, VA, VB and VC are observed, each of which consists of quadrupole split seven lines and marked by the red, green and blue arrows. (b) and (c) The angle dependences of peak frequencies with $H_{\rm{ext}}$ applied in the $ca'$- and $a'b'$-planes. In (c), the data with blue open squares (red solid dots) are assigned to V$^{5+}$(1) and V$^{5+}$(3) (V$^{5+}$(2) and V$^{5+}$(4)) sites. The solid lines show the fitting described in the text.} 
\end{figure}

Figure \ref{fig:300K_V_spc} (a) shows the $^{51}$V NMR spectrum from V$^{5+}$ sites obtained at 300 K with the magnetic field of 6.615 T applied in the $ca'$-plane. Because of the glide symmetry with respect to the $ca'$-plane which exchanges V$^{5+}$(1) and V$^{5+}$(3) sites each other, these two sites should give the identical NMR spectrum. Since other two sites are not related by such a symmetry that leaves the field direction invariant, we expect three sets of NMR spectra, each of which consists of quadrupole split seven lines. This is indeed the case as marked by the red, green and blue arrows in Fig.~\ref{fig:300K_V_spc} (a) and labeled as VA, VB, and VC, respectively. The peak frequencies of the quadrupole split lines for each set are plotted in Fig.~\ref{fig:300K_V_spc}(b) with the same colors. Full width of half maximum (FWHM) of the center line is quite narrow, about 5 kHz for each set, indicating high quality of the sample.  Since the intensity of VC is twice as strong as VA and VB, VC can be assigned to V$^{5+}$(1) and V$^{5+}$(3) sites.  However, other two sites cannot be assigned uniquely. Therefore we consider two cases; in case 1, VA (VB) is assigned to V$^{5+}$(2) (V$^{5+}$(4)) and vice versa in case 2. When the field is parallel to $c$, all sites are indistinguishable due to $C_4$ symmetry.

Figure \ref{fig:300K_V_spc} (c) shows the resonance frequencies for the magnetic field in the $a'b'$-plane. In this case, the field direction remains unchanged by $C_2$ along the $c$-axis, which exchanges V$^{5+}$(1) and V$^{5+}$(3), as well as V$^{5+}$(2) and V$^{5+}$(4). Therefore, the spectra consists of two sets, VD (blue open squares) and VE (red solid dots). To be compatible with the assignment in Fig.~\ref{fig:300K_V_spc}(b) for ${\bf{H}}_{\rm{ext}} \parallel a'$, VD (VE) must be assigned to V$^{5+}$(1) and V$^{5+}$(3) sites (V$^{5+}$(2) and V$^{5+}$(4) sites). 

\begin{figure}[t]
\includegraphics[width=8.5cm]{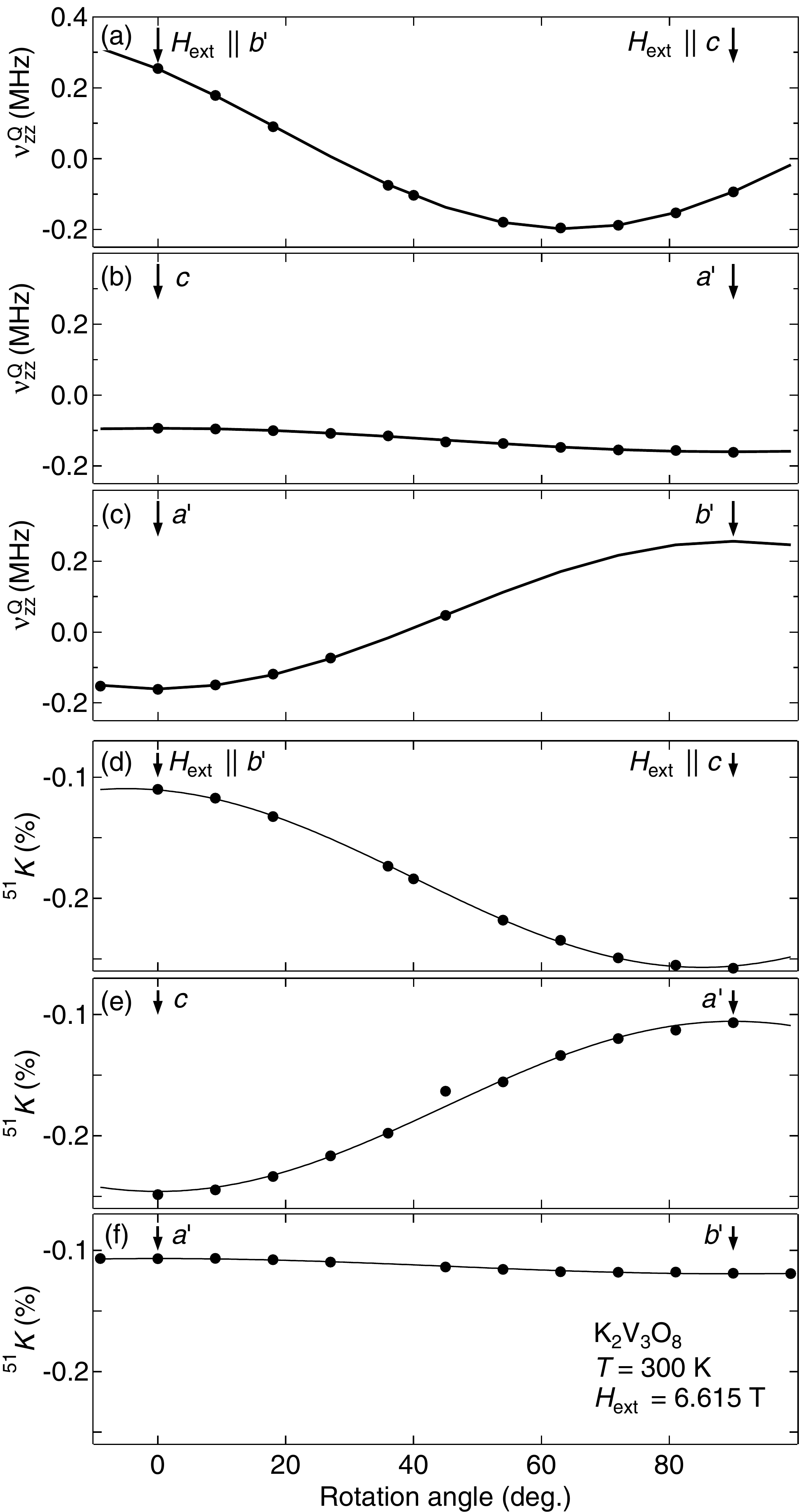}
\caption{\label{fig:300K_nuq_K} Angle dependences of $\nu^Q_{zz}$ and $K$ at V$^{5+}$(1) site with the field $H_{\rm{ext}}$ = 6.615 T rotated in the (a), (d) $b'c'$-, (b), (e) $ca'$-, and (c), (f) $a'b'$-planes at $T$ = 300 K.  The solid lines show the fitting described in the text.} 
\end{figure}

From the NMR frequencies of each set, the values of the quadrupole splitting $\nu^Q_{zz}$ and the shift $K$ are determined by Eqs.~(\ref{centerline}), (\ref{Qcoupling}), and (\ref{shift}). Using the data for VC and VD in Fig.~\ref{fig:300K_V_spc}(b) and (c), $\nu^Q_{zz}$ and $K$ for V$^{5+}$(1) site at $T$ = 300 K are obtained as a function of the field direction in the $ca'$-, and $a'b'$-planes and plotted in Fig.~\ref{fig:300K_nuq_K}(b), (c), (e), and (f). Because of the $C_4$ symmetry, the NMR spectra of the V$^{5+}$(1) site with fields in the $b'c$-planes should be identical to the spectra of the V$^{5+}$(4) site with fields in the $ca'$-plane. Therefore, $\nu^Q_{zz}$ and $K$ for V$^{5+}$(1) site with fields in the $b'c$-planes are obtained from the VB data in Fig.~\ref{fig:300K_V_spc}(b), assuming case 1, and plotted in Fig.~\ref{fig:300K_nuq_K}(a) and (d).  

From these data, one can determine all components of the shift and the quadrupole coupling tensors as follows. For the fields in the $ca'$-plane, for example, the angle dependence of $K$ is expressed by using Eqs.~(\ref{shift}) and (\ref{Ktensor}) with ${\bf{h}} = (\sin\theta, 0, \cos\theta)$ as
\begin{equation}
K = \frac{K_{cc}+K_{a'a'}}{2} + \frac{K_{cc}-K_{a'a'}}{2}\cos 2\theta ,
\end{equation}
which is used to fit the data in Fig.~\ref{fig:300K_nuq_K}(e). Similarly, the data in Fig.~\ref{fig:300K_nuq_K}(d) and (e) can be fit to the functions, $(K_{b'b'}+K_{cc})/2 + (K_{b'b'}-K_{cc})\cos 2\theta /2 + (K_{b'c}+K_{cb'})\sin 2\theta /2$ and $(K_{a'a'}+K_{b'b'})/2 + (K_{a'a'}-K_{b'b'})\cos 2\theta /2$, respectively. Note that the off-diagonal elements appear only in the symmetric form $K_{b'c}+K_{cb'}$. The same analysis can be applied to the quadrupole coupling tensor only by replacing the components of ${\bf{K}}$ by those of ${\bm{\nu}}^{\rm{Q}}$ in the fitting functions. The results of the fitting are shown by the lines in Figs.~\ref{fig:300K_nuq_K} and \ref{fig:300K_V_spc}. The components of ${\bf{K}}$ and ${\bm{\nu}}^{\rm{Q}}$ for V$^{5+}$(1) site at 300 K are determined as 
\begin{eqnarray}
{{\bf{K}}}_1=\left(\begin{array}{ccc}
-0.11 & 0 & 0 \\
0 & -0.11 & -0.01\\
0 & -0.01 & -0.25
\end{array}\right) \%
\label{eq:K_300},\\
{{\bm{\nu}}^{\rm{Q}}}_1=\left(\begin{array}{ccc}
-0.08& 0 & 0 \\
0 & 0.13 & -0.11\\
0 & -0.11 & -0.05
\end{array}\right) {\rm{MHz}}
\label{eq:nu_300}
\end{eqnarray}
for case 1. We assumed ${\bf{K}}_1$ to be symmetric.

\begin{figure}[t]
\includegraphics[width=6cm]{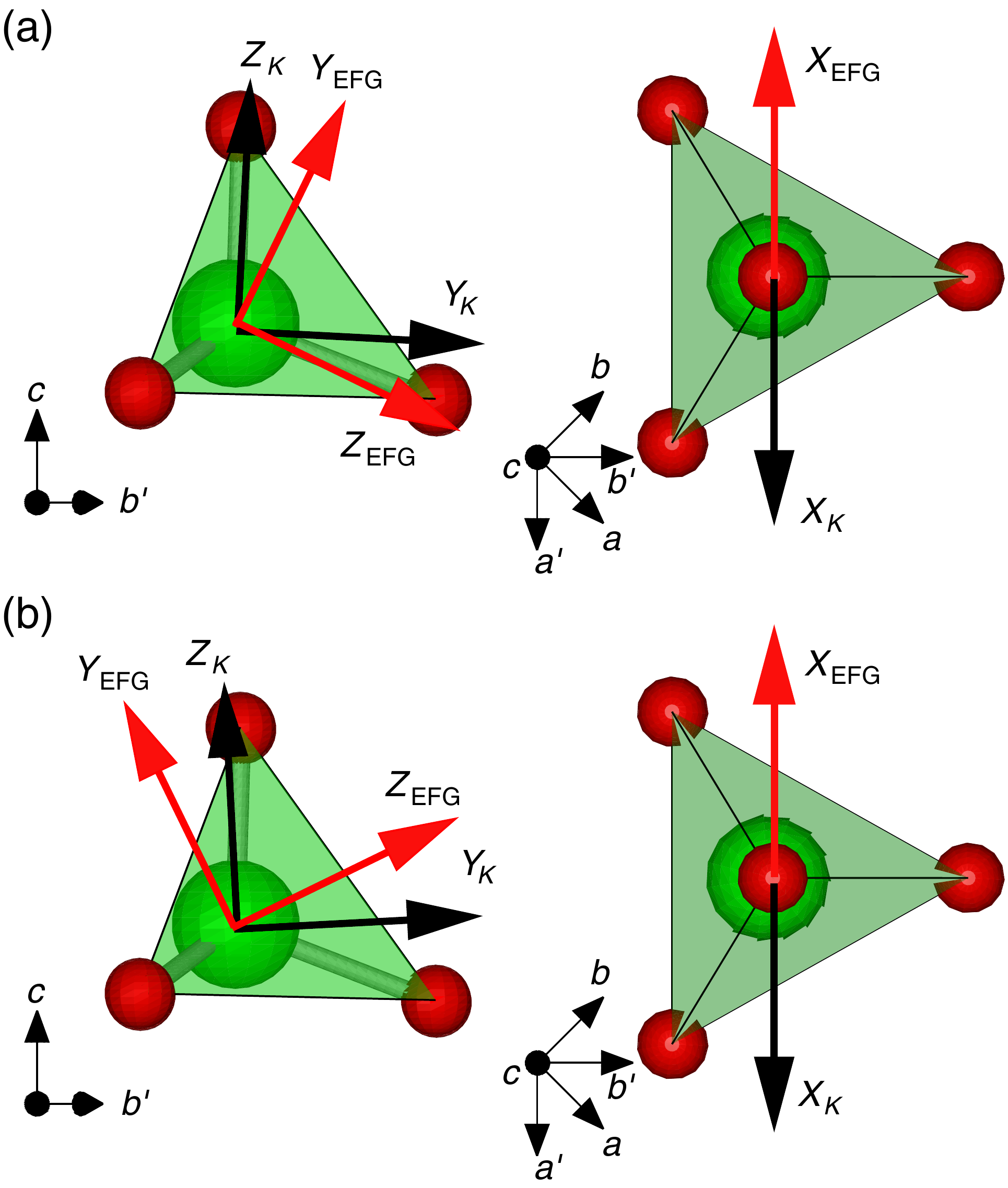}
\caption{\label{fig:V5_local_axis} (Color Online) Principal axes of the ${\bf{K}}$ and ${\bm{\nu}}^{\rm{Q}}$ tensors for V$^{5+}$(1) site for (a) case 1 and (b) case 2.} 
\end{figure}

\begin{table}[t]
\caption{\label{tab:table1} Principal values of the ${\bf{K}}$ and ${\bm{\nu}}^{\rm{Q}}$ tensors for the V$^{5+}$(1) site of K$_2$V$_3$O$_8$ at 300 K. $\alpha_{\rm{K}}$ ($\alpha_{\rm{Q}}$) denotes the tilting angle of the $Z$-axis of ${\bf{K}}$ tensor (the $Y$-axis of ${\bm{\nu}}^{\rm{Q}}$ tensor) from $c$ to $b'$ in case 1. For case 2, $\alpha_{\rm{K}}$ and $\alpha_{\rm{Q}}$ change sign, while other parameters remain the same.}
\begin{ruledtabular}
\begin{tabular}{ccccc}
 & $K$ (\%) & $\alpha_{\rm{K}}$ & $\nu^{\rm{Q}}$ (MHz) & $\alpha_{\rm{Q}}$ \\
\hline
$X$ & $-$0.11 &  ---         & $-$0.08 & ---  \\
$Y$ & $-$0.11 &  ---          & $-$0.10 & 26$^{\circ}$ \\
$Z$ & $-$0.25 & 3$^{\circ}$ & 0.18 & --- \\
\end{tabular}
\end{ruledtabular}
\end{table}

By diagonalizing these tensors, the principal values and the corresponding principal axes are determined as shown in Table \ref{tab:table1} and Fig.~\ref{fig:V5_local_axis} for both case 1 and case 2. Here, $Z$ ($X$) denotes the principal axis corresponding to the largest (smallest) absolute principal value. As indicated in Eq.~(\ref{Ktensor}), $a'$ is one of the principal axes due to mirror symmetry.  It turns out that $a'$ is the $X$-axis for both ${\bf{K}}$ and ${\bm{\nu}}^{\rm{Q}}$. The other principal axes are in the $b'c$-plane. The $Z$-axis of ${\bf{K}}$ ($Y$-axis of ${\bm{\nu}}^{\rm{Q}}$) are tilted from $c$ to $b'$ by an angle $\alpha_{\rm{K}}$ ($\alpha_{\rm{Q}}$), which changes sign between case 1 and case 2, while the principal values remain the same for the two cases. Since $\alpha_{\rm{K}}$ is rather small, the choice of the two cases does not make significant difference in the following analysis on the spin structure in the antiferromagnetic state and the spin-lattice relaxation rate. Therefore, we assume case 1 in the following analysis and discussion. 

\begin{figure}[t]
\includegraphics[width=8.5cm]{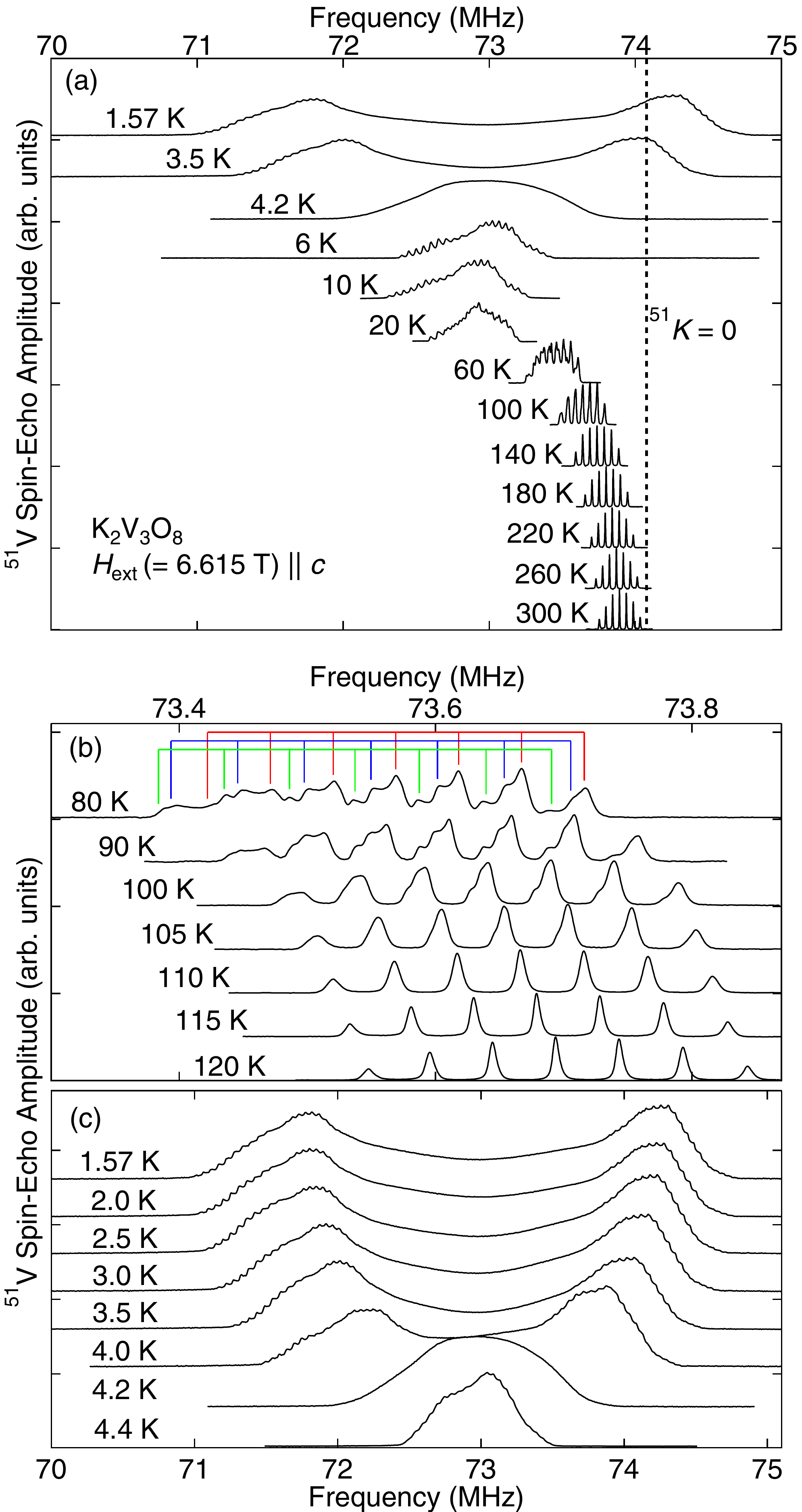}
\caption{\label{fig:spc_c} (Color online) (a) Temperature dependence of the $^{51}$V NMR spectra for $H_{\rm{ext}}=6.615$ T along the $c$-axis. Detailed $T$ dependences of the spectra below the  structural transition at $T_{\rm{S}}$ = 115~K and the magnetic transition at $T_{\rm{N}} \sim 4$~K are shown in (b) and (c).  The red, blue and green lines indicate appearance of three sets of NMR lines.} 
\end{figure}

Next, we discuss temperature dependence of the NMR spectrum. Figure~\ref{fig:spc_c} (a) shows the variation of the NMR spectra with temperatures for ${\bf{H}}_{\rm{ext}}||c$. Upon cooling, the spectrum shifts to lower frequency and gradually splits below $T_{\rm{S}}=115$~K into three sets with the intensity ratio of approximately 1 : 2 : 3 as shown in Fig. \ref{fig:spc_c} (b). The continuous evolution of the spectral shape suggests a second order transition, consistent with the X-ray scattering measurement.\cite{Chakoumakos}  The lines get broadened with further decreasing temperature. Due to combined effects of threefold splitting and broadening of lines, it is not possible to resolve individual peaks below 30 K. Below $T_{\rm{N}} \sim 4$~K, the spectral shape turns into a double peak structure as shown in Fig. \ref{fig:spc_c}(c). This is ascribed to the appearance of internal magnetic field generated by spontaneous magnetic moments. 

\begin{figure}[t]
\includegraphics[width=8cm]{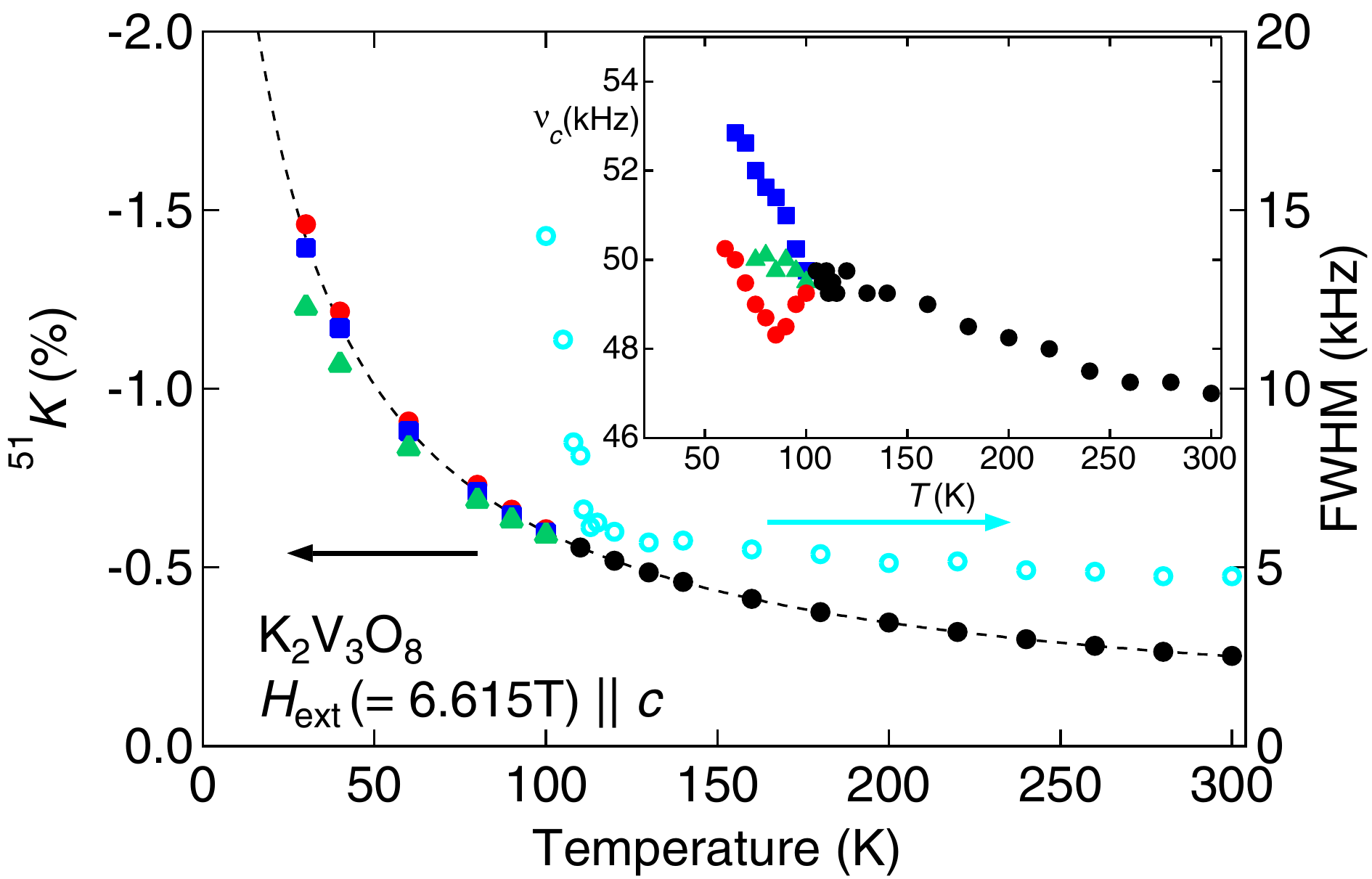}
\caption{\label{fig:K_temp2} (Color online) Temperature dependence of the shift $K_{cc}$ and FWHM of the center line at $H_{\rm{ext}}=6.615$~T along the $c$-axis.  The dashed curve shows the fitting to the Curie-Weiss formula $K=K_0+C/(T-\theta)$. The inset shows $T$ dependence of the quadrupole splitting $\nu^{\rm{Q}}_{cc}$.} 
\end{figure}

\begin{figure}[t]
\includegraphics[width=8cm]{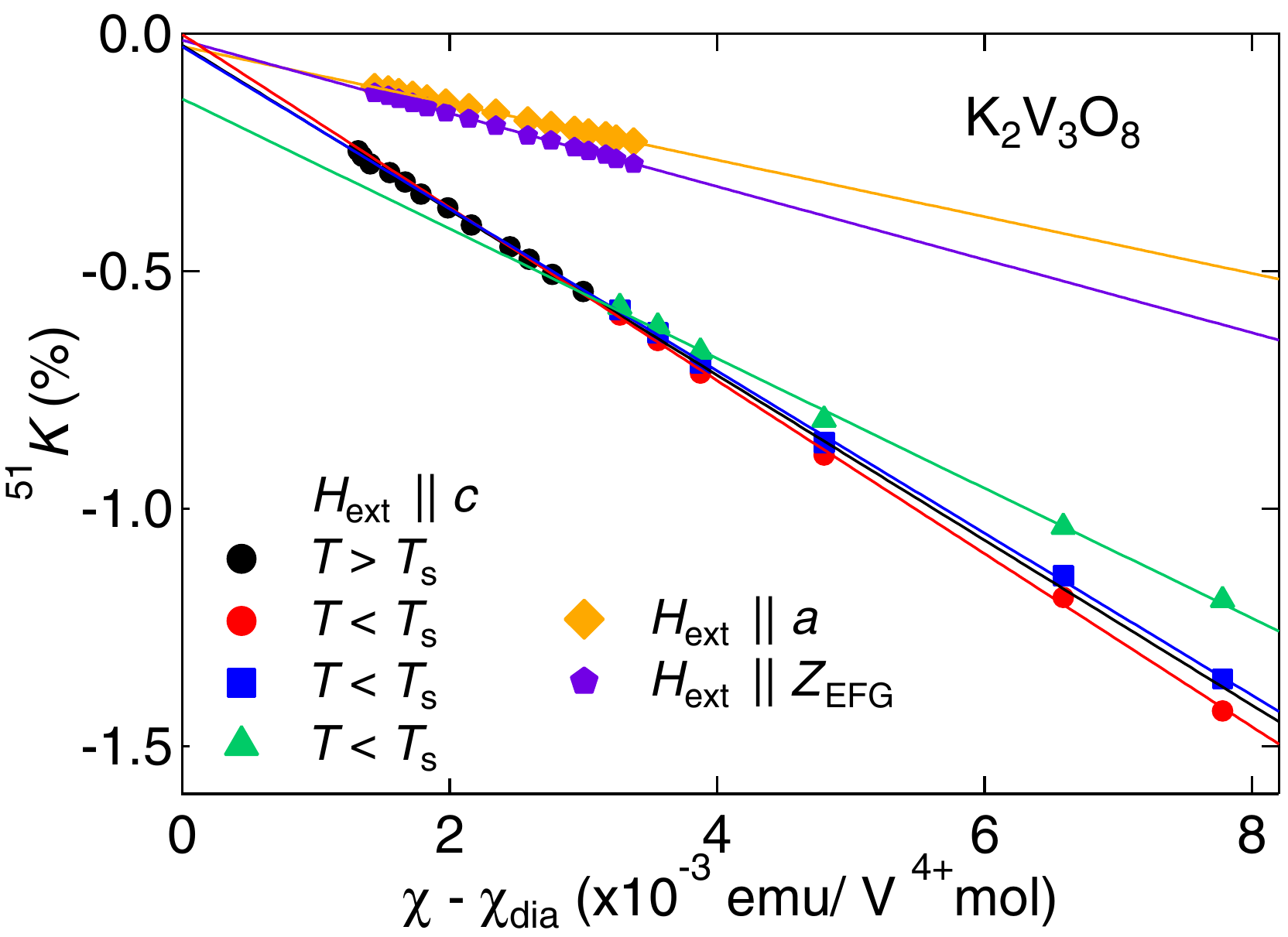}
\caption{\label{fig:K_chi} (Color online) The shift $K$ at V$^{5+}$(1) sites plotted against the magnetic susceptibility $\chi$ for three different field directions: $c$, $a$ and the $Z_{\rm EFG}$.} 
\end{figure} 

Figure \ref{fig:K_temp2} shows the $T$ dependences of $K_{cc}$, the shift for ${\bf{H}}_{\rm{ext}}||c$. The data can be fit to the Curie-Weiss law, $K=K_0+C/(T-\theta)$ in the range 110 K $\leq $ $T$ $\leq $ 300 K with the parameter values, $K_0=-0.047$ $\%$, $C=-65$ K$^{-1}$ and $\theta=-17$ K.  This value of $\theta$ is almost the same as the Weiss temperature of the susceptibility, $\theta$ = $-$16 K.\cite{Liu}  In Fig.~\ref{fig:K_chi}, $K$ at V$^{5+}$(1) sites measured above $T_{\rm{S}}$ is plotted against the magnetic susceptibility $\chi$ for three different directions of $H_{\rm{ext}}$. Each of these $K$ vs. $\chi$ plots can be fit to a straight line consistent with Eq.~(\ref{Kchi}). Since the anisotropy of $\chi$ is negligibly small in the paramagnetic phase, the hyperfine coupling tensor ${\bf{A}}$ can be determined from the slope of the $K$ vs. $\chi$ plots as
\begin{eqnarray}{\label{coupling}}
{{\bf{A}}}=\left(\begin{array}{ccc}
-0.33 & 0 & 0\\
0 & -0.33 & -0.03\\
0 & -0.03 & -0.97
\end{array}\right){\rm{T}}/\mu_{\rm{B}}.
\end{eqnarray}
Here we assumed that the principal axes of $K$ shown in Table \ref{tab:table1} does not change with temperature. The coupling tensor ${\bf{A}}$ is the sum of contributions from the classical dipolar interaction and the transferred hyperfine interaction caused by covalent bonding effects, ${\bf{A}} = {\bf{A}}^{\rm{dip}} +  {\bf{A}}^{\rm{tr}}$. The dipole contribution can be calculated by summation over lattice points as 
\begin{eqnarray}{\label{dip}}
{{\bf{A}}}^{\rm{dip}}=\left(\begin{array}{ccc}
0.058 & 0 & 0\\
0 & -0.019 & 0.0003\\
0 & 0.0003 & -0.040
\end{array}\right){\rm{T}}/\mu_{\rm{B}}, 
\end{eqnarray}
leaving the transferred hyperfine coupling tensor as
\begin{eqnarray}{\label{demag}}
{{\bf{A}}}^{\rm{tr}}=\left(\begin{array}{ccc}
-0.39 & 0 & 0\\
0 &-0.31& -0.03\\
0 &-0.03& -0.93
\end{array}\right){\rm{T}}/\mu_{\rm{B}}.
\end{eqnarray}
Note that ${\bf{A}}^{\rm{tr}}$ is an order of magnitude larger than ${\bf{A}}^{\rm{dip}}$. 

The structural transition at 115 K causes splitting of the spectrum, while no visible anomaly appears in the magnetic susceptibility.\cite{JChoi}  This indicates that the hyperfine coupling tensor must be affected by the transition. Indeed, the $K_{cc}$ versus $\chi$ plots for the split peaks in the range 60 K $\leq T\leq $ 110 K below $T_{\rm{S}}$ shown in Fig. \ref{fig:K_chi} give different values of the coupling constant $A_{cc}$, $-1.02$, $-0.95$, $-0.76$T/$\mu_{\rm{B}}$, compared to the single value of $-0.97$ T/$\mu_{\rm{B}}$ above $T_{\rm{S}}$.  The quadrupole coupling is also affected by the structural transition. As shown in the inset of Fig. \ref{fig:K_temp2}, $\nu^{\rm{Q}}_{cc}$ for the split peaks exhibit widely different $T$ dependences and rapidly changing even below 60 K . This indicates that the lattice distortion keeps growing down to low temperatures. Because of these unusual spectral features, we are unable to resolve individual peaks below 60 K.   

\subsection{Magnetically ordered phase}{\label{AFphase}}

\begin{figure}[t]
\includegraphics[width=8.5cm]{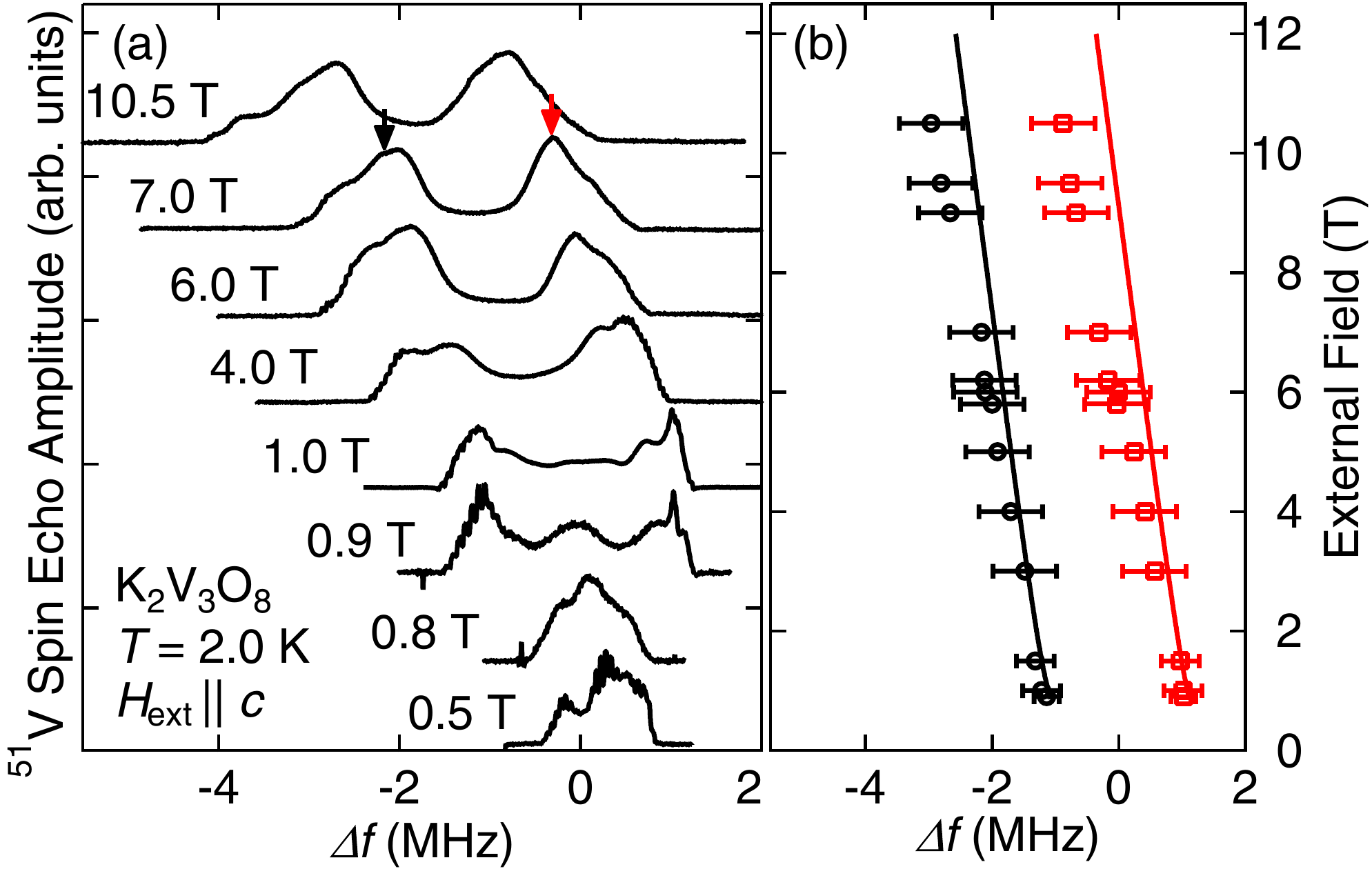}
\caption{\label{fig:spc_2K_field} (Color online)  (a) External field dependence of $^{51}$V NMR spectra obtained at 2 K for $H_{\rm{ext}}||c$.   $\Delta f$ is defined as the frequency shift measured from the reference frequency $\Delta f=f-\gamma H_{\rm{ext}}$. (b) External field dependence of the frequencies of the two peaks indicated by the arrows in (a) above 0.9 T.  The solid curves show the calculation based on the model described in the text.} 
\end{figure}

\begin{figure}[t]
\includegraphics[width=7cm]{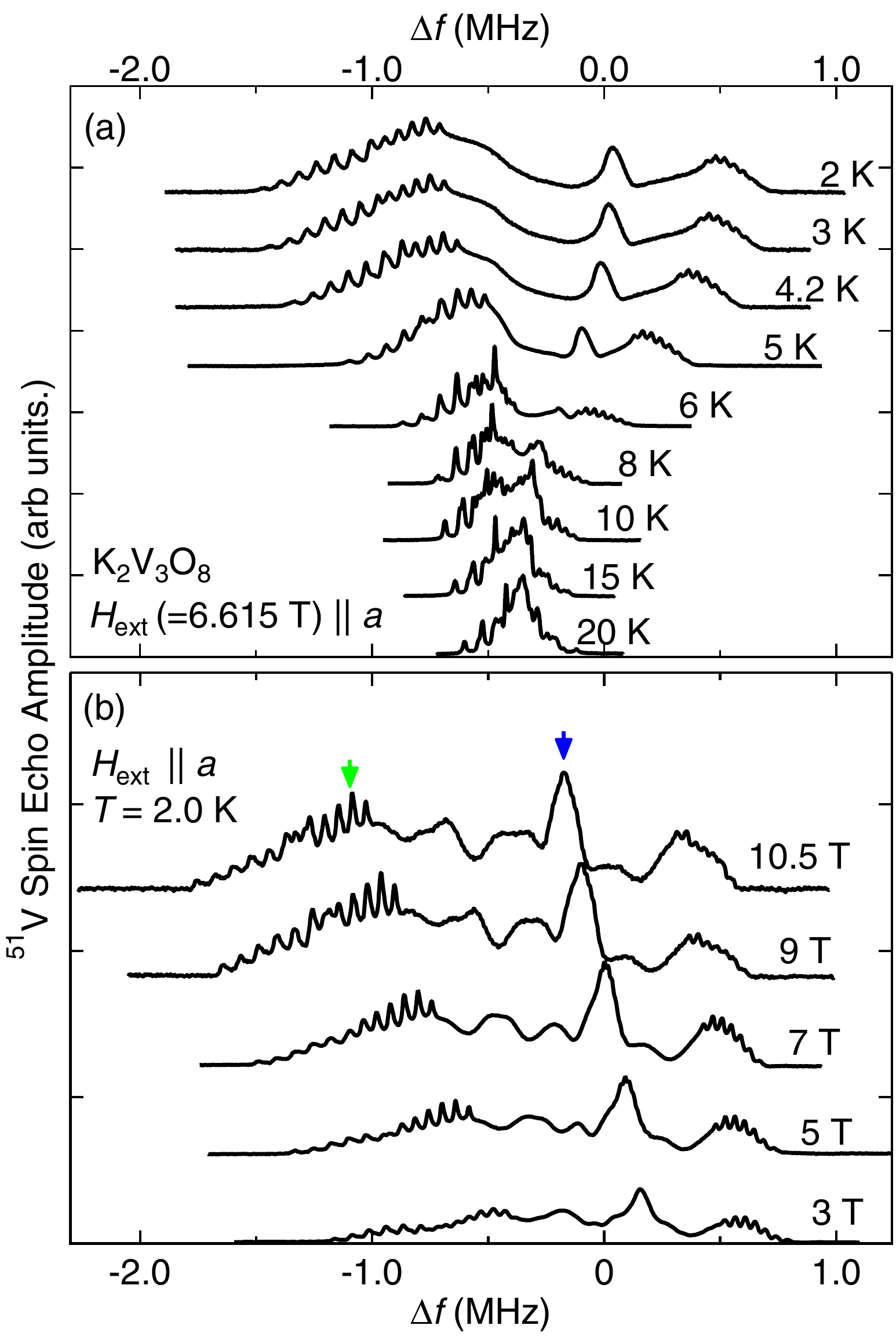}
\caption{\label{fig:spc_2K_field_a} (Color online) (a) Temperature dependence of the $^{51}$V NMR spectra for $H_{\rm{ext}}$ = 6.615 T along the $a$-axis. (b) External field dependence of $^{51}$V NMR spectra obtained at 2 K for $H_{\rm{ext}}||a$.  Two arrows indicate the spectral positions at which $1/T_1$ was measured.} 
\end{figure}

We next focus on the $^{51}$V NMR spectra for $T<T_{\rm{N}} \sim$~4K.  As shown in Fig. \ref{fig:spc_c} (c), broad spectra with double peak structure were observed for $H_{\rm{ext}}$ = 6.615~T along the $c$-axis. This spectral shape does not change much with magnetic field in the range 0.9 $\leq H_{\rm{ext}}\leq $ 10.5~T as shown in Fig.~\ref{fig:spc_2K_field}(a). At lower fields, however, a drastic change occurs from double peak to single peak structure with significant narrowing. This is ascribed to the spin flop transition at 0.85~T\cite{Lumsden1} as we discuss in section \ref{AFstructure}. The broad spectral shape with double peaks at the edges is often associated with an incommensurate spin structure. However, in our case, the broad feature appears already above $T_{\rm{N}}$ due to structural complexity. Therefore, we cannot make definite conclusion whether the magnetic structure is commensurate or incommensurate.      

When the field is applied along the $a$-axis, we observed broad spectra with more complex spectral shape as shown in Fig. \ref{fig:spc_2K_field_a}.  However, this spectral shape persists even above $T_{\rm{N}}$, therefore, it is mainly governed by the distributions of hyperfine fields and nuclear quadrupole couplings caused by the structural transition.  For ${\bf{H}}_{\rm{ext}}||a$ the spectral shape may change below 0.65~T due to continuous reorientation of the spin structure.  However, we were not able to detect NMR signals below 0.65~T due to low signal intensity and very short spin-echo decay time $T_2$. 

\subsection{Spin lattice relaxation rate}{\label{1/T1}}

\begin{figure}[t]
\includegraphics[width=8cm]{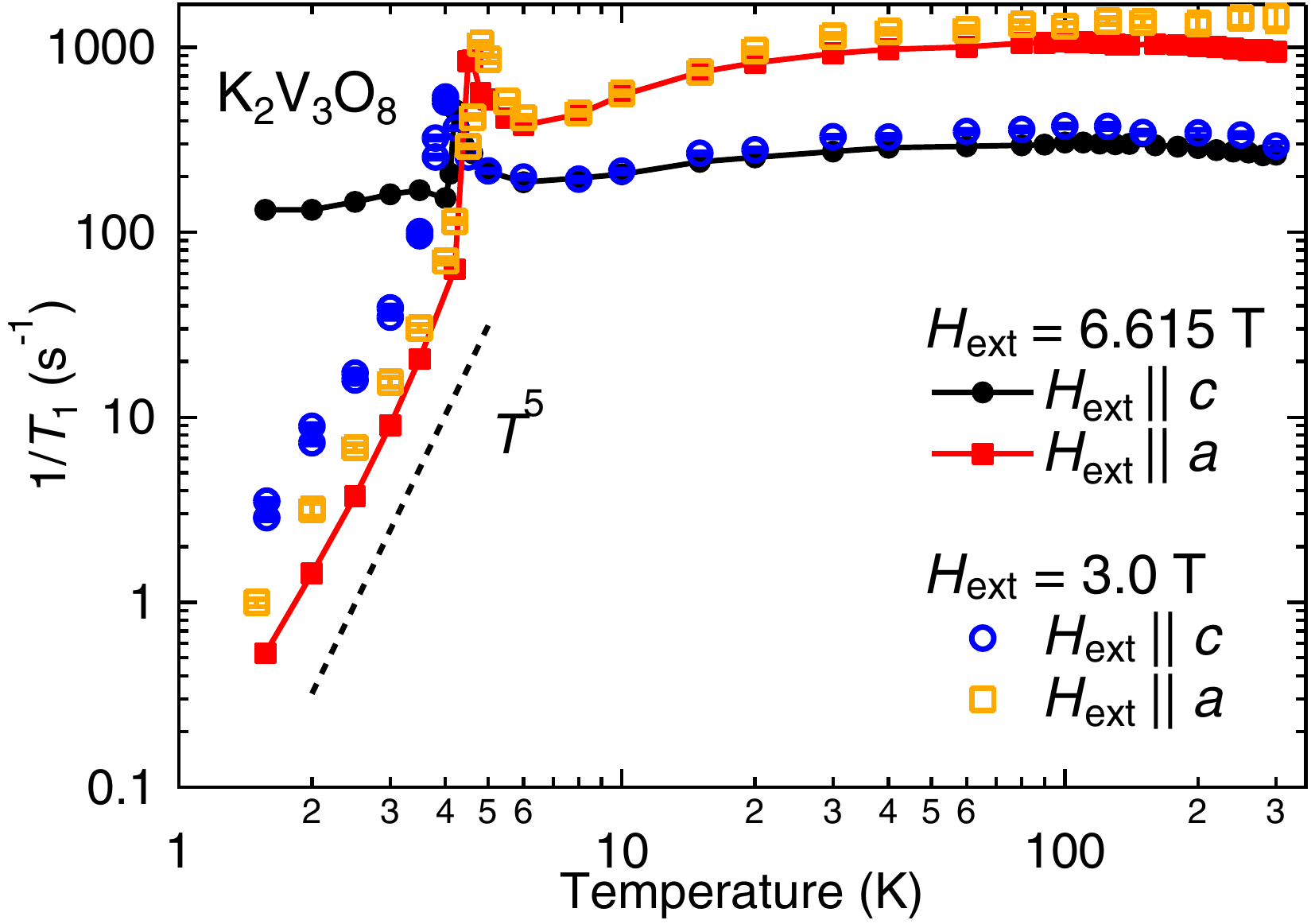}
\caption{\label{fig:T1_temp} (Color online) Temperature dependence of the spin-lattice relaxation rate $1/T_1$ measured in the external fields $H_{\rm{ext}}$ = 3.0 and 6.615~T applied along the $a$- and $c$-axes.} 
\end{figure}

\begin{figure}[t]
\includegraphics[width=8cm]{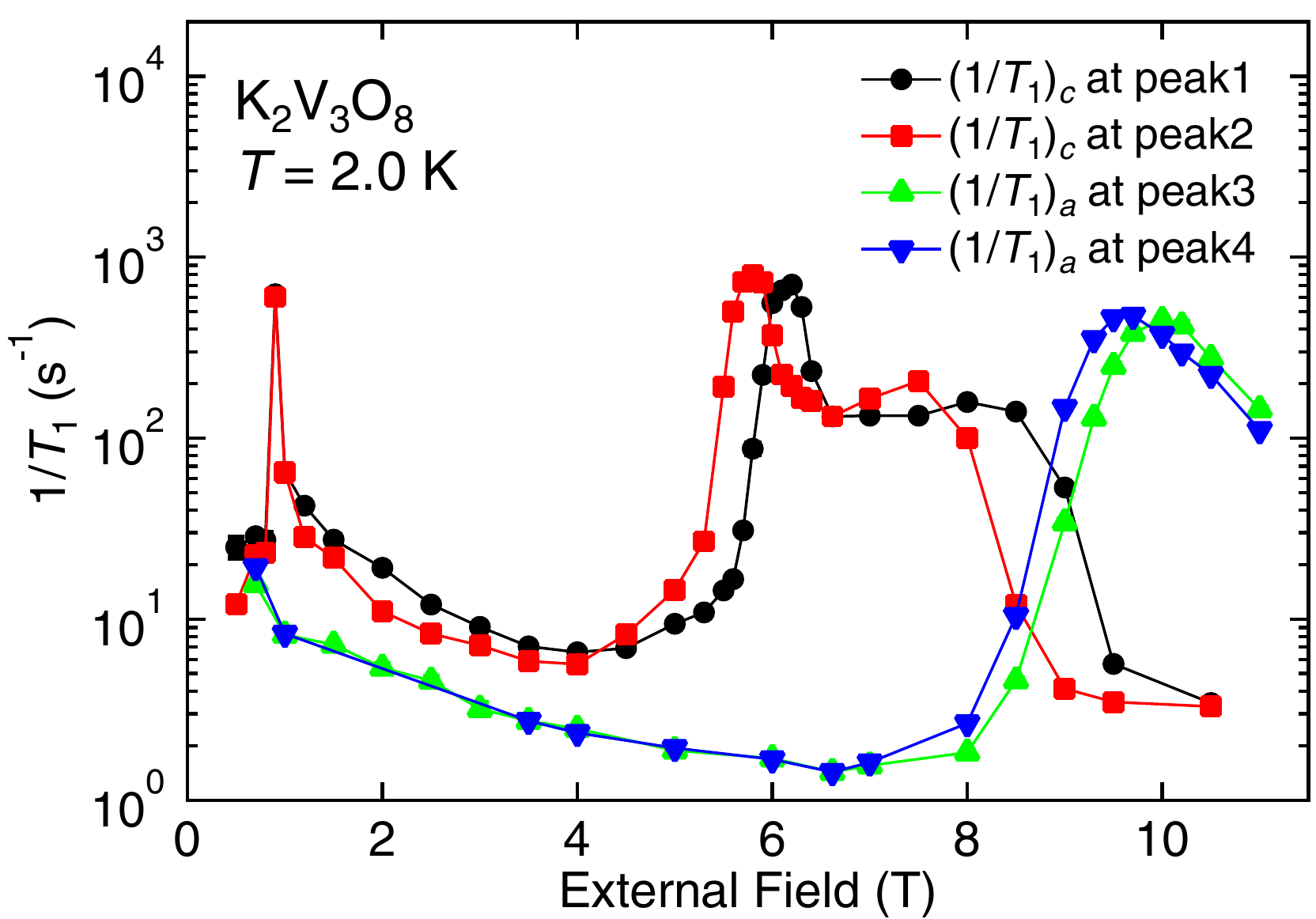}
\caption{\label{fig:T1_2K} External field dependence of the spin-lattice relaxation rate $1/T_1$ at 2 K measured at the peak frequencies of the spectra marked in Figs. \ref{fig:spc_2K_field} and \ref{fig:spc_2K_field_a}(b). The color of the plots indicates correspondence to the peak in Figs. \ref{fig:spc_2K_field} and \ref{fig:spc_2K_field_a}(b) marked with the same color. } 
\end{figure}

Figure~\ref{fig:T1_temp} shows the $T$ dependence of the spin-lattice relaxation rates measured for $H_{\rm{ext}}$ = 6.615~T applied along the $a$- and $c$-axes.  Above 30 K, $(1/T_1)$ are nearly independent of temperature for both field directions. Such a behavior is typical for magnetic insulators in the temperature range sufficiently higher than $T_{\rm{N}}$. In spite of the line splitting at $T_{\rm{S}}$, $(1/T_1)$ does not show any anomaly at the structural transition.   

Upon cooling below 30 K, $(1/T_1)_a$ for ${\bf{H}}_{\rm{ext}}||a$ decreases gradually down to 6~K, below which it shows a divergent behavior toward $T_{\rm{N}}\sim$ 4 K. Below $T_{\rm{N}}$, $(1/T_1)_a$ decreases steeply, following approximately the $T^5$ dependence. Similar behavior is observed also for $H_{\rm{ext}}$ = 3.0~T.  These  power law behaviors are typical $T$ dependence for an antiferromagnet whose nuclear spin relaxation rate is governed by two or three magnon scattering processes.\cite{Beeman}  
For ${\bf{H}}_{\rm{ext}}||c$, the $T$ dependence of $(1/T_1)_c$ is similar to that of $(1/T_1)_a$ above $T_{\rm{N}}$.  However, the $T$ dependence of $(1/T_1)_c$ below $T_{\rm{N}}$ changes drastically with magnetic field. While $(1/T_1)_c$ at $H_{\rm{ext}}$ = 6.615~T depends only very weakly on $T$ down to 1.5~K, it shows a steep decrease at $H_{\rm{ext}}$ = 3~T similar to the behavior of $(1/T_1)_a$.  Such a weak $T$ dependence of $1/T_1$ suggests appearance of low energy spin fluctuations in a high magnetic field.  

In order to understand this puzzling field dependence, we have performed detailed $1/T_1$ measurements over a wide range of magnetic fields at 2~K as shown in Fig.~\ref{fig:T1_2K}. Since the spectra are quite broad below $T_{\rm{N}}$, we measured $1/T_1$ at two frequencies for each field direction; the two peaks (peak 1 and 2) marked by the black and red arrows in Fig.~\ref{fig:spc_2K_field} for ${\bf{H}}_{\rm{ext}}||c$ and the similar peak structures (peak 3 and 4) marked by the green and blue arrows in Fig.~\ref{fig:spc_2K_field_a} for ${\bf{H}}_{\rm{ext}}||a$. For both field directions, similar field dependence of $1/T_1$ is observed at different frequencies. With increasing field for ${\bf{H}}_{\rm{ext}}||c$, $(1/T_1)_c$ first exhibits a very sharp peak at the spin flop transition (0.85~T). Such a behavior has been observed also in other antiferromagnets.\cite{Paquette}  In the field range $2<H_{\rm{ext}}<5$~T, $(1/T_1)_c$ keeps low values $\sim$10 s$^{-1}$. However, $(1/T_1)_c$ shows a steep enhancement over two orders of magnitude in a narrow field window above 5~T and keeps high values up to $\sim$8~T, above which it is suppressed steeply again to the values below 10 s$^{-1}$. On the other hand, $(1/T_1)_a$ for ${\bf{H}}_{\rm{ext}}||a$ keeps low values up to $\sim$8~T but get enhanced over nearly three orders of magnitude within the field range 8 $\sim$ 10~T, in a qualitatively similar manner to the behavior of $(1/T_1)_c$ above 5~T. We propose in section \ref{discussion} that such an anomalous field dependence of $1/T_1$ can be caused by cross relaxation effects between nuclei on non-magnetic V$^{5+}$ sites and those on magnetic V$^{4+}$ sites.

\section{DISCUSSION}{\label{discussion}}
\subsection{Lattice distortion}{\label{distortion}}
Below $T_{\rm{S}}$ = 115 K, the $^{51}$V NMR spectrum for ${\bf{H}}_{\rm{ext}}||c$ splits into three sets as shown in Fig.~\ref{fig:spc_c}. This splitting is most likely associated with the superlattice formation with the wave vector $\frac{1}{3} \langle 110 \rangle ^{*} + \frac{1}{2} c^{*}$ detected by X-ray diffraction.\cite{Chakoumakos} In the following, we propose a possible mechanism which relates these two phenomena. 

\begin{figure}[t]
\includegraphics[width=7cm]{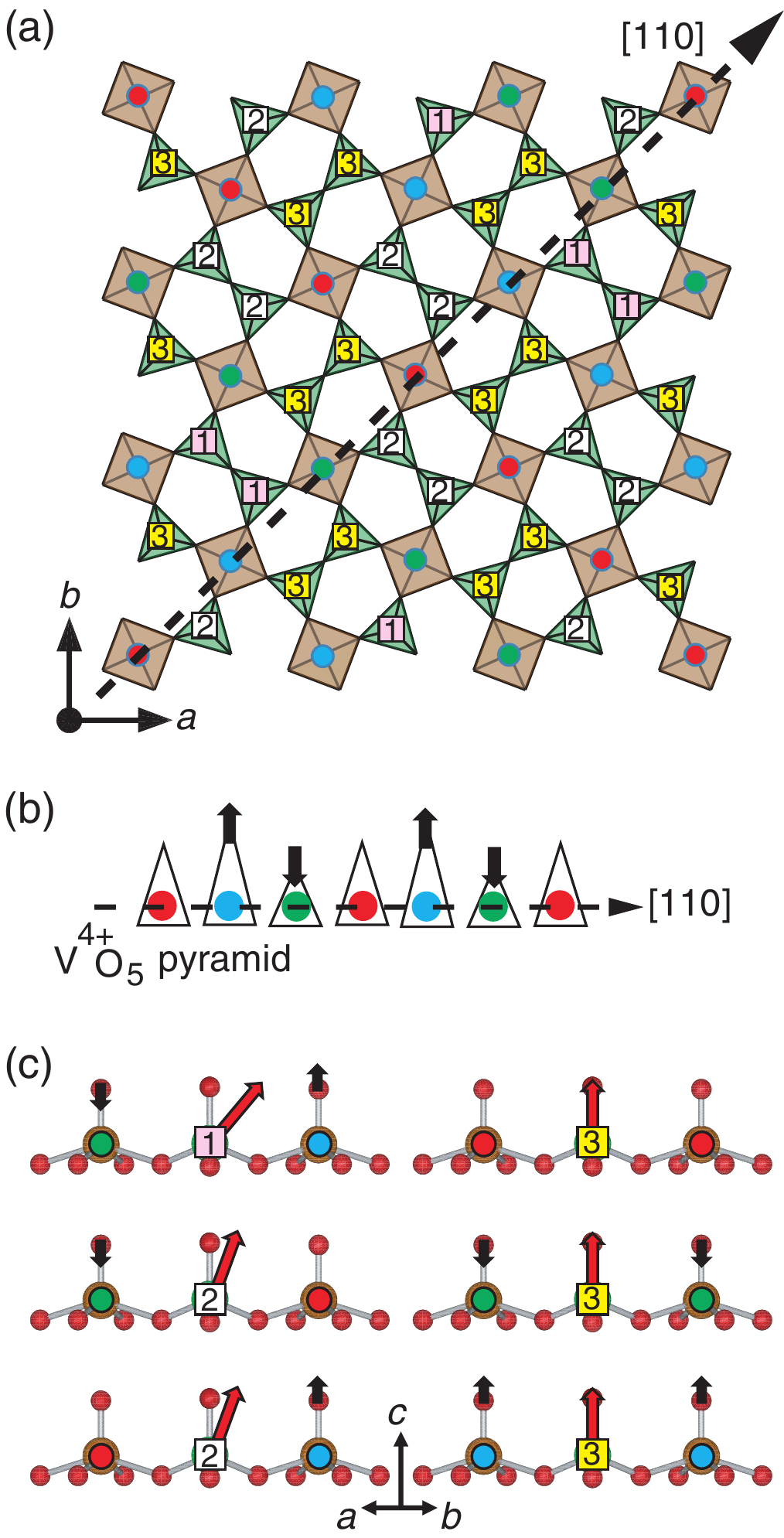}
\caption{\label{fig:distortion} (Color online) (a) The crystal structure with the 3$\times$3$\times$2 supperlattice below 115~K projected on the $ab$ plane.  (b) Proposed modulation of V$^{4+}$O$_5$ pyramids along [110].  (c) Three types of local distortion surrounding V$^{5+}$O$_4$ tetrahedra connected to two V$^{4+}$O$_5$ pyramids.  Black arrows on the V$^{4+}$O$_5$ pyramids denote distortion of the apical oxygens, while red arrows in the V$^{5+}$O$_4$ tetrahedra indicate tilted principal axes $Z$ for the Knight shift tensor.} 
\end{figure}

Infrared and Raman spectroscopy measurements revealed local lattice distortion in the V$^{4+}$O$_5$ pyramid,\cite{JChoi, Rai, KChoi} in particular, the stretching phonon mode for the V$^{4+}$-O$^{\rm{ap}}$ (apical oxygen) bond exhibits splitting at 115 K.  Then it is likely that the V$^{4+}$-O$^{\rm{ap}}$ bond length has the same periodic modulation as the supertattice below $T_{\rm{S}}$ as illustrated in Fig.~\ref{fig:distortion}(a) and (b). There are three types of V$^{4+}$O$_5$ pyramids indicated by different colors, blue, red, and green, with different V$^{4+}$-O$^{\rm{ap}}$ bond length, longest in the blue pyramids and shortest in the green pyramids. Because each V$^{5+}$ sites is chemically bonded to two nearest neighbor V$^{4+}$O$_5$ pyramids, six types of of V$^{5+}$ sites can be distinguished according to the combination of the types of the nearest neighbor V$^{4+}$O$_5$ pyramids: (1) blue-blue, (2) red-red, (3) green-green, (4) blue-red, (5) red-green, and (6) greed-blue. In the first three cases, the V$^{5+}$ sites are connected to the same type of pyramid and the local mirror symmetry is preserved. We then expect only a minor change in the hyperfine coupling constant. In the other three cases, the V$^{5+}$ sites are connected to different types of pyramids, which breaks the mirror symmetry and results in the tilting of the principal $Z$-axis of the $K$ tensor from the mirror plane. Such a drastic change in symmetry is likely to cause substantial change in the hyperfine coupling. Moreover, since the degree of tilting should be larger for case (6) than cases (4) and (5), the intensity ratio of 1 : 2 : 3 for the split NMR lines can be reproduced by this model.

\subsection{Spin structure in the magnetically ordered phase}{\label{AFstructure}}

The NMR spectra below $T_{\rm{N}}$ show broad two peak structure with finite spectral intensity extending over the entire frequency range. Such a continuous distribution of hyperfine field in a single crystal is usually associated with an incommensurate spin structure. In our case, however, quasi-continuous distribution of the hyperfine field already appears above $T_{\rm{N}}$ due to structural modulation with a large supercell. Therefore, the distribution of the hyperfine field below $T_{\rm{N}}$ is, at least partially, due to distribution of the hyperfine coupling constant and the modulation of the antiferromagnetic moments caused by the structural distortion. Since the precise crystal structure below $T_{\rm{S}}$ is not known, we are unable to analyze such effects quantitatively. On the other hand, we show in the following that gross features of the field dependence of the spectum for ${\bf{H}}_{\rm{ext}}||c$ can be simply understood based on the spin structure proposed by the neutron scattering experiments\cite{Lumsden1} built on the undistorted high temperature structure.

As mentioned in section \ref{para}, the hyperfine field at the V$^{5+}$ sites comes dominantly from the short range transferred hyperfine interaction rather than the long range dipolar interaction. Therefore, we only consider contribution from the two nearest neighbor V$^{4+}$ moments in the summation of Eq.~(\ref{hfcoupling}) for the V$^{5+}$(1) sites in Fig.~\ref{fig:crystal}, ${\bf{H}}_1^{\rm{hf}}={\bf{A}}_{\rm{1A}}\cdot{\bm{\mu}}_{\rm{A}}+{\bf{A}}_{\rm{1B}}\cdot{\bm{\mu}}_{\rm{B}}$, where 
${\bf{A}}_{\rm{1A}}$ and ${\bf{A}}_{\rm{1B}}$ are the hyperfine coupling tensors to the moment on the V$^{4+}$(A) and  V$^{4+}$(B) site in Fig.~\ref{fig:crystal} and include both the transferred hyperfine and the dipolar contributions. These are expressed as  
\begin{eqnarray}
{\bf{A}}_{\rm{1A}}=\left(\begin{array}{ccc}
-0.16 & \delta & \epsilon\\
\delta & -0.17 & -0.02\\
\epsilon & -0.02 & -0.49
\end{array}\right) {\rm{T/\mu_B}} 
\label{eq:one},\\
{\bf{A}}_{\rm{1B}}=\left(\begin{array}{ccc}
-0.16 & -\delta & -\epsilon\\
-\delta & -0.17 & -0.02\\
-\epsilon & -0.02 & -0.49
\end{array}\right) {\rm{T/\mu_B}} .
\label{eq:two}
\end{eqnarray}
Note that the off-diagonal components $\delta$ and $\epsilon$ are generally non-zero because the individual bonds V$^{5+}$(1) - V$^{4+}$(A) and V$^{5+}$(1) - V$^{4+}$(B) are not on the mirror plane. The values of these components change sign between ${\bf{A}}_{\rm{1A}}$ and ${\bf{A}}_{\rm{1B}}$ due to the mirror symmetry, therefore, their contribution vanish when the moments are uniform as in the paramagnetic state. The hyperfine fields at the other V$^{5+}$ sites can be obtained by using the $C_4$ transformation.   

The neutron diffraction experiments\cite{Lumsden1} have proposed a simple N\'{e}el order for ${\bf{H}}_{\rm{ext}}||c$ below the spin-flop transition field (0.85~T), in which the corner and face-center moments within the basal plane are aligned along the $c$ axis and antiparallel each other; thus ${\bm{\mu}}_{\rm{A}}=(0,0,M)$ and ${\bm{\mu}}_{\rm{B}}=(0,0,-M)$. The hyperfine fields at the four V$^{5+}$ sites in a unit cell are then expressed as
\begin{align}
\begin{aligned}{\label{Hn1}}
{\bf{H}}_1^{\rm{hf}}&=2\epsilon M(1, 0, 0), \\
{\bf{H}}_2^{\rm{hf}}&=2\epsilon M(0, 1, 0), \\
{\bf{H}}_3^{\rm{hf}}&=2\epsilon M(-1, 0, 0), \\
{\bf{H}}_4^{\rm{hf}}&=2\epsilon M(0, -1, 0)
\end{aligned}
\end{align}
in the $a'b'c$ coordinate frame. For all sites, ${\bf{H}}^{\rm{hf}}$ is perpendicular to ${\bf{H}}_{\rm{ext}}||c$ and has the same magnitude, therefore, no splitting is expected below $T_{\rm{N}}$. This is consistent with the observed spectra for $H_{\rm{ext}}$($||c$) = 0.5 and 0.8 T at 2.0 K shown in Fig.~\ref{fig:spc_2K_field}(a). These spectra have relatively narrow width and the center of gravity is shifted from $\gamma H_{\rm{ext}}$ by about 0.2 and 0.1~MHz, respectively. This shift is expressed as $\gamma |{\bf{H}}_{\rm{ext}} + {\bf{H}}_{\rm{hf}}| - \gamma H_{\rm{ext}}$. By taking $M$ = 0.7 $\mu_{\rm{B}}$\cite{Lumsden1},  the observed shift for $H_{\rm{ext}}$($||c$) = 0.5 and 0.8~T can be reproduced by assuming $|\epsilon|$ = 0.10~T/$\mu_{\rm{B}}$.  

The field along the $c$ axis above 0.85~T causes the antiferromanetic moments to flop from $c$ to $a$.\cite{Lumsden1} With further increasing field, the uniform magnetization are induced along the $c$ axis. Therefore, the sublattice moments are expressed as ${\bm{\mu}}_{\rm{A}} = M(u/\sqrt{2}, u/\sqrt{2}, v)$ and ${\bm{\mu}}_{\rm{B}} = M(-u/\sqrt{2}, -u/\sqrt{2}, v)$ with $u^2+v^2=1$. The hyperfine fields are then given as  
\begin{align}
\begin{aligned}{\label{Hn2}}
{\bf{H}}_1^{\rm{hf}}&=\sqrt{2}Mu(\delta, \delta, \epsilon)+2Mv(0,-0.02,-0.49), \\ 
{\bf{H}}_2^{\rm{hf}}&=\sqrt{2}Mu(-\delta, -\delta, \epsilon)+2Mv(0.02,0,-0.49), \\
{\bf{H}}_3^{\rm{hf}}&=\sqrt{2}Mu(\delta, \delta, -\epsilon)+2Mv(0,0.02,-0.49), \\
{\bf{H}}_4^{\rm{hf}}&=\sqrt{2}Mu(-\delta, -\delta, -\epsilon)+2Mv(-0.02,0,-0.49).
\end{aligned}
\end{align}
The first term from the in plane AF moments has a staggered $c$ component and produces two peaks of the spectrum. The second term due to the field-induced uniform magnetization shifts the entire spectrum to lower frequency. The field induced moment is determined from the magnetization data in Ref.\onlinecite{Rai} as $v = \kappa  H$ with $\kappa = 0.016$~T$^{-1}$ for $H_{\rm{ext}}<$~10~T.  Assuming again $|\epsilon|$ = 0.10~T/$\mu_{\rm{B}}$ and $M$ = 0.7 $\mu_{\rm{B}}$, the frequencies of the two peaks are calculated by $f_i=\gamma|{\bf{H}}_{\rm{ext}}+{\bf{H}}_i^{\rm{hf}}|$ ($i$ = 1-4) and displayed in Fig.~\ref{fig:spc_2K_field}(b) by the solid lines. The calculated frequency shifts agree well with the observed positions of the broad peaks at low fields below 4~T, although systematic deviation develops at higher fields.

The quantitative agreement between the calculation and the experimental spectra indicates that the simple two-sublattice spin structure is not fundamentally affected by the structural modulation.  Such a magnetic structure was also reported in another oxide with similar quasi-two dimensional lattice, Ca$_2$CoSi$_2$O$_7$.\cite{Soda}  In these systems, the small lattice modulations appear to have negligible effects on the exchange interactions and the spin anisotropy which determine the spin structure.\cite{Zaliznyak, Lumsden2}  In K$_2$V$_3$O$_8$, the structural transition has no effects on  $1/T_1$ as mentioned in section \ref{1/T1}, indicating that the exchange interaction indeed remains unchanged by the structural transition.  However, we should note that possibility of an incommensurate spin structure cannot be ruled out because of the broad nature of the spectra. It is thus difficult to discuss presence or absence of the modulated magnetic phase as proposed by Bogdanov $et$ $al$.\cite{Bogdanov} based on the NMR results.  

\subsection{1/$T_1$ above $T_{\rm{N}}$}

The nuclear spin-lattice relaxation rate in magnetic insulators is determined by the time correlation function of spins coupled to nuclei. When temperature is sufficiently higher than the energy scale of exchange interaction, dynamics of individual spins can be modeled by random Gaussian fluctuations under the influence of exchange coupled neighboring spins and  $1/T_1$ is expressed as\cite{Moriya1, Moriya2} 
\begin{equation}{\label{eq_T1}}
\begin{split}
1/T_1=&\frac{\sqrt{\pi/2}S(S+1)}{3\hbar^2\omega_e} \\
&\times\sum_{i} \left[(1-h_{X}^2)A_{iXX}^2+(1-h_{Y}^2)A_{iYY}^2\right. \\
&\left. +(1-h_{Z}^2)A_{iZZ}^2 \right]. 
\end{split}
\end{equation}
Here, the exchange frequency $\omega_e$ is defined by $\omega_e^2=2J^2pS(S+1)/3\hbar^2$ with the exchange interaction $J$ and the number of neighboring spins $p$ ($p = 4$ for K$_2$V$_3$O$_8$).  $A_{iXX}$, $A_{iYY}$, and $A_{iZZ}$ are principal values of hyperfine coupling tensors, and $h_{X}$, $h_{Y}$, and $h_{Z}$ are the components of the unit vector along ${\bf{H}}_{\rm{ext}}$ in the $XYZ$ coordinate frame.  We consider contribution from two nearest neighbor spins, Eqs.~(\ref{eq:one}) and (\ref{eq:two}), in the sum of Eq.~(\ref{eq_T1}).

Taking the values $\delta=\epsilon=-1$ kOe/$\mu_{\rm{B}}$, as discussed in section \ref{AFstructure}, the principal values of the hyperfine coupling tensor are obtained as ($A_{iXX}$, $A_{iYY}$, $A_{iZZ}$) = ($-0.06$, $-0.24$, $-0.52$) T/$\mu_{\rm{B}}$.  Then, using Eq. (\ref{eq_T1}) with these principal values, $S=1/2$, and $|J|=12.0$ K determined from the spin wave spectrum obtained by inelastic neutron scattering measurements \cite{Lumsden2}, 
we can estimate $(1/T_1)_a$ and $(1/T_1)_c$ as 475 s$^{-1}$ and 1615 s$^{-1}$. 
These values are larger than the experimental values at 300~K and 6.615~T, $(1/T_1)_c=260$ s$^{-1}$ and $(1/T_1)_a =950$ s$^{-1}$ by a factor 1.7 - 1.8 but discrepancy is not significant.. 
It should be stressed that there is no anomaly in the $T$ dependence of $1/T_1$ at $T_{\rm{S}} \sim$ 115 K, indicating that exchange interactions are not largely affected by the lattice distortion. 

\subsection{Enhancement of $1/T_1$}
In contrast to the weak field dependence of the NMR spectra, $1/T_1$ shows anomalously strong dependence on $H_{\rm{ext}}$ (Fig.~\ref{fig:T1_2K}). The most distinctive feature is the sharply defined field range 5 $<H_{\rm{ext}}<$ 9~T, in which $(1/T_1)_c$ for ${\bf{H}}_{\rm{ext}}||c$ is enhanced over more than an order of magnitude. Neither magnetization nor NMR spectra shows such anomaly. This motivated us to look into the cross relaxation phenomena caused by mutual coupling between two distinct spin systems, one with slow relaxation and the other with fast relaxation. Their Larmor frequencies are generally different, therefore, exchange of energy between two spin systems are inhibited. However, if they can coincide by adjusting the external field, the relaxation rate of the slow system will be largely enhanced by exchanging Zeeman energy with the fast system.  

\begin{figure}[t]
\includegraphics[width=8cm]{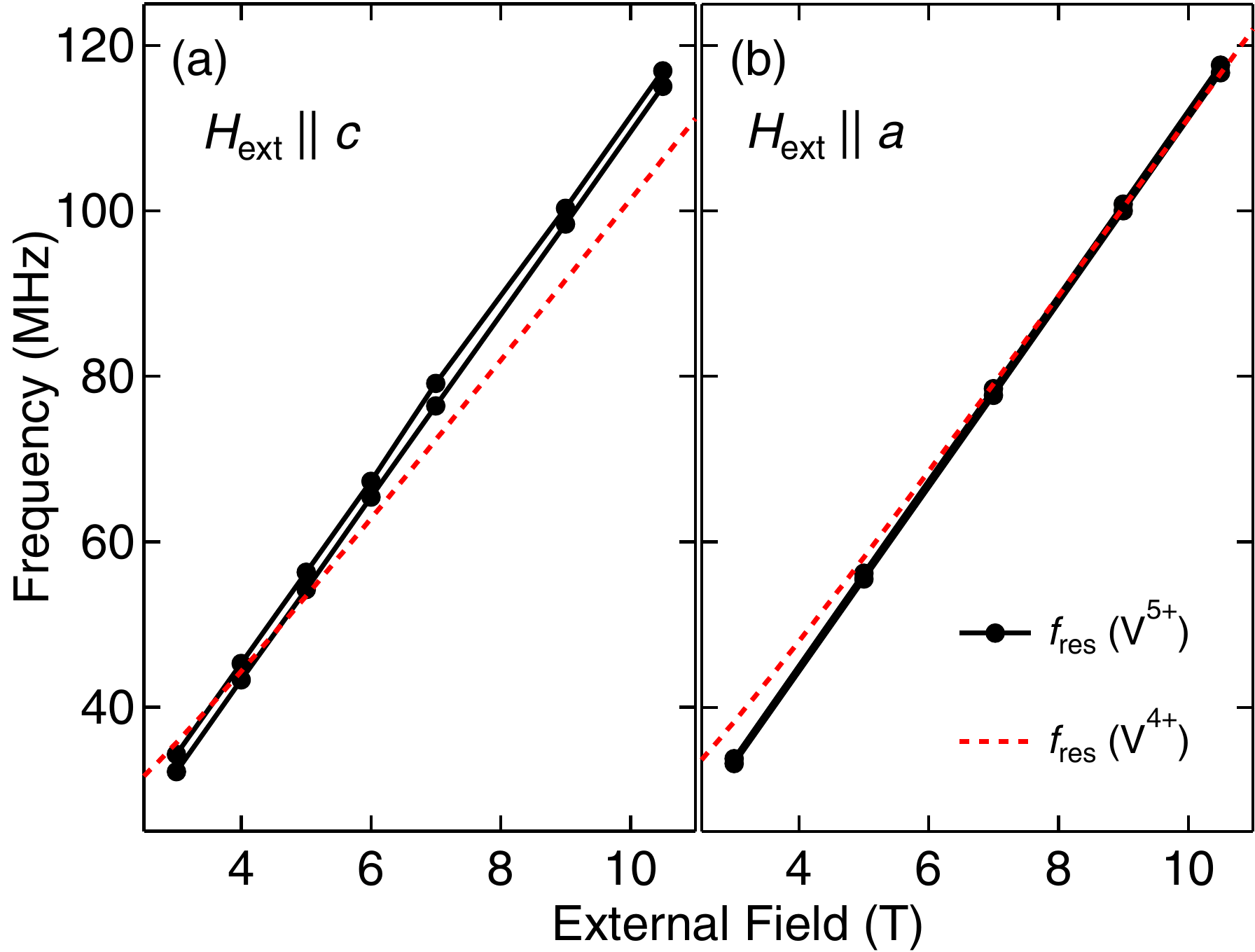}
\caption{\label{fig:V4_V5} (Color online) The dashed red line indicate the calculated NMR frequency of V$^{4+}$ sites as a function of the external magnetic field along the (a) $c$-axis and (b) $a$-axis as described in the text. The peak frequencies of the V$^{5+}$ NMR spectra marked in Figs.~\ref{fig:spc_2K_field} and \ref{fig:spc_2K_field_a}(b), at which $1/T_1$ is measured, are also plotted by solid circles. }
\end{figure}

In our case, the slow and the fast system can be $^{51}$V nuclei on V$^{5+}$ and V$^{4+}$ sites, respectively. Although we are unable to observe NMR signal from V$^{4+}$ sites likely due to extremely short spin-echo decay time ($T_2$), their Larmor frequency can become close to that of V$^{5+}$ sites. The V$^{4+}$ sites in K$_2$V$_3$O$_8$ form V$^{4+}$O$_5$ pyramids with one 3$d$ electron occupying the $d_{xy}$ orbital. The hyperfine field from onsite 3$d$ electrons is usually an order of magnitude larger than the transferred hyperfine field from neighboring sites. In order to estimate the hyperfine field at the V$^{4+}$ sites, we refer to the reported results on $\alpha$-NaV$_2$O$_5$, which also contains V$^{4+}$O$_5$ pyramids with one 3$d$ electron in the $d_{xy}$ orbital.\cite{Ohama} Ohama $et$ $al$. performed $^{51}$V NMR experiments on the V$^{4+}$ sites in $\alpha$-NaV$_2$O$_5$ and found an almost uniaxial hyperfine coupling tensor with the principal values reported ($-$2.0, $-$2.7, $-$10.2)~T/$\mu_{\rm{B}}$.  

By assuming the hyperfine coupling tensor ($-$2.5, $-$2.5, $-$10.0)~T/$\mu_{\rm{B}}$ for K$_2$V$_3$O$_8$ with uniaxisal symmetry along the $c$-axis, the nuclear Larmor frequency of V$^{4+}$ sites is calculated as a function of field and shown by the red dashed line in Fig.~\ref{fig:V4_V5}(a) for ${\bf{H}}_{\rm{ext}}||c$ and (b) for ${\bf{H}}_{\rm{ext}}||a$. We employed the same model for the evolution of spin structure with field as discussed in section \ref{AFstructure} for ${\bf{H}}_{\rm{ext}}||c$. For ${\bf{H}}_{\rm{ext}}||a$, we assumed spin reorientation to $b$-axis below 1~T and growth of the uniform magnetization along the $a$-axis at higher fields.\cite{Lumsden1} The frequencies of the two peaks of the V$^{5+}$ NMR spectra, at which $1/T_1$ is measured, are also plotted by solid circles. Comparing the NMR frequencies of V$^{4+}$ and V$^{5+}$ sites, they indeed coincide near $H_{\rm{ext}} \sim 4$~T for ${\bf{H}}_{\rm{ext}}||c$ and near $H_{\rm{ext}} \sim 9$~T for ${\bf{H}}_{\rm{ext}}||a$ in reasonable agreement with the peak of $(1/T_1)_c$ at 6~T and peak of $(1/T_1)_a$ at 10~T (Fig.~\ref{fig:V4_V5}). The slight disagreement can be attributed to ambiguity in the hyperfine field from onsite 3$d$ electrons since it depends on chemical environments. The peak fields of $(1/T_1)_c$ and $(1/T_1)_a$ can be reproduced exactly if we choose the hyperfine coupling as ($-$3.5, $-$3.5, $-$8.1)~T/$\mu_{\rm{B}}$. Thus the enhancement of $1/T_1$ at 2 K is likely to be associated with cross relaxation between nuclei on V$^{4+}$ and V$^{5+}$ sites.  

\begin{figure}[t]
\includegraphics[width=8.5cm]{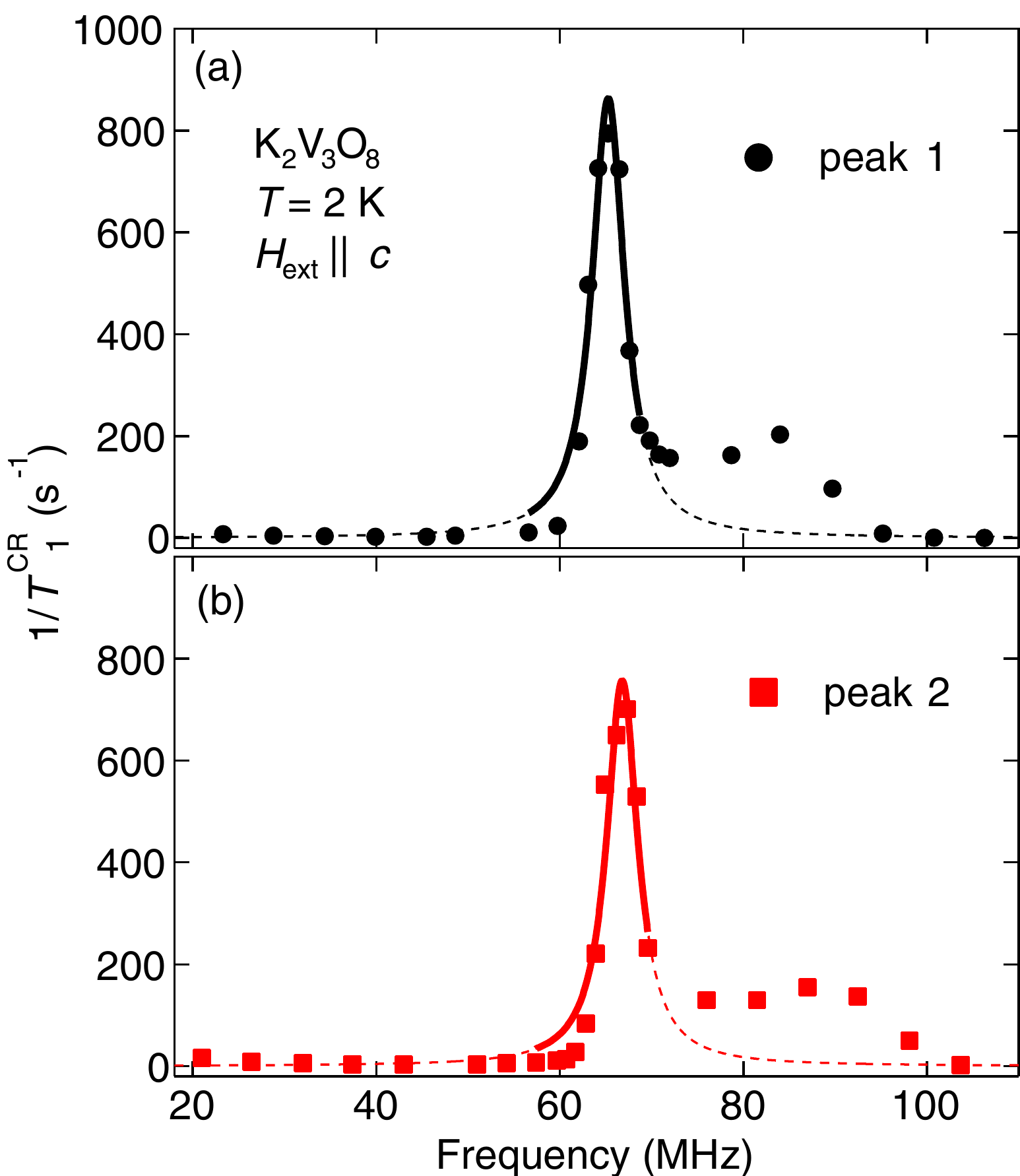}
\caption{\label{fig:T1_freq_analysis} (Color online) Frequency dependence of the cross relaxation rate defined as $1/T_1^{\rm{CR}}=1/T_1-1/T_1(H_{\rm{ext}}=10.5$ T), where $1/T_1$ is measured at (a) peak 1 and (b) peak 2 for ${\bf{H}}_{\rm{ext}}||c$ at 2~K.  The lines show the results of fitting to Eq.~(\ref{CRfit}) with the fitting range indicated by the thick part.} 
\end{figure}  

In order to examine the validity of cross relaxation mechanism quantitatively, we follow the analysis presented in Refs.[\onlinecite{Abragam, Chudo, Tokunaga, Tokunaga2}]. The coupling between nuclei on V$^{4+}$ and V$^{5+}$ sites relevant to the spin-lattice relaxation of V$^{5+}$ nuclei can be written as 
\begin{equation}{\label{spin-spin}}
\begin{split}
{\mathcal{H}}_{\rm{cr}}& =\sum_{j, k} \alpha_{jk} \left( I^{4+}_{+j} I^{5+}_{-k} + I^{4+}_{-j} I^{5+}_{+k} \right) \\
                       & + \sum_{j, k} \beta_{jk}  I^{4+}_{zj} \left( I^{5+}_{-k} + I^{5+}_{+k} \right),
\end{split}
\end{equation}
where $I^{4+}_{\pm j}$ ($I^{4+}_{zj}$) is the component of the nuclear spin perpendicular (parallel) to the time-averaged local field at the $j$-th V$^{4+}$ site and $I^{5+}_{\pm k}$ ($I^{5+}_{zk}$) is similarly defined for V$^{5+}$ sites. This coupling enables the relaxation processes of nuclear magnetization at V$^{5+}$ sites, that is the transitions between different eigenstates of $I^{5+}_{zk}$, caused by fluctuations of either $xy$- or $z$-component of the V$^{4+}$ nuclear spins. By assuming Lorentzian frequency spectra for fluctuations of V$^{4+}$ nuclear spins, which is centered at $\omega$ = 0 for the $z$-component and at $\omega_{4+}$, the Larmor frequency of the V$^{4+}$ nuclei, for the $xy$-component, the cross relaxation rate $1/T_1^{\rm{CR}}$ of the V$^{5+}$ nuclei is given as a function of the NMR angular frequency $\omega$ of V$^{5+}$ nuclei as\cite{Chudo, Abragam, Tokunaga, Tokunaga2}
\begin{equation}{\label{CR}}
\frac{1}{T_1^{\rm{CR}}}=\frac{\braket{\Delta\omega^2}_{\alpha} \tau_{\alpha}}{1+(\omega-\omega_{4+})^2 \tau_{\alpha}^2}+\frac{\braket{\Delta\omega^2}_{\beta} \tau_{\beta}}{1+\omega^2 \tau_{\beta}^2},
\end{equation}
where $\tau_{\alpha}$ ($\tau_{\beta}$) is the correlation time of $z$ ($xy$) component of the V$^{4+}$ nuclear spins and $\braket{\Delta\omega^2}_{\alpha, \beta}$ are the respective contribution to the second moment of the local field at the V$^{5+}$ site. 
 
The cross relaxation rate $1/T_1^{\rm{CR}}$ for ${\bf{H}}_{\rm{ext}}||c$ is obtained experimentally from the data of $1/T_1$ in Fig.~\ref{fig:T1_2K} by subtracting the value of $1/T_1$ at the highest field of 10.5~T, where the cross relaxation effect is absent. The results are shown in Fig.~\ref{fig:T1_freq_analysis}(a) for peak 1 and (b) for peak 2. The sharp peak of $1/T_1^{\rm{CR}}$ at 66~MHz suggests that the first term is dominant in Eq.~(\ref{CR}). Keeping only the first term, we fit the data of $1/T_1^{\rm{CR}}$ in Fig.~\ref{fig:T1_freq_analysis} to the following function, 
\begin{equation}{\label{CRfit}}
\frac{1}{T_1^{\rm{CR}}}=\frac{\braket{\Delta\omega^2}_{\alpha} \tau_{\alpha}}{1+4\pi^2 (f - f_{4+})^2 \tau_{\alpha}^2}. 
\end{equation}
The values of the fitting parameters $\sqrt{\braket{\Delta\omega^2}_{\alpha}}$, $\tau_{\alpha}$, and $f_{4+}$ are obtained as 
$\sqrt{\braket{\Delta\omega^2}_{\alpha}} = (1.08 \pm 0.04) \times 10^{5}$ s$^{-1}$, $\tau_{\alpha} = (7.5 \pm 0.9) \times 10^{-8}$ s, $f_{4+} = 65.3\pm0.2$ MHz for peak 1 and $\sqrt{\braket{\Delta\omega^2}_{\alpha}} = (0.98 \pm 0.05) \times 10^{5}$ s$^{-1}$, $\tau_{\alpha} = (7.8 \pm 1.1) \times 10^{-8}$ s, $f_{4+} = 66.8\pm0.2$ MHz for peak 2. The correlation time $\tau_{\alpha}$ provides the upper limit of the spin-echo decay rate $1/T_2$ of V$^{4+}$ nuclear spins. The very short $\tau_{\alpha} \sim$ 0.1 $\mu$s is, therefore, consistent with the absence of observable V$^{4+}$ NMR signal. However, we also note that the peak-width of $1/T_1^{\rm{CR}}$ in Fig.~\ref{fig:T1_freq_analysis} is assumed to be entirely dynamic in our analysis. There should be some static distribution of $f_{4+}$ due to inhomogeneity of antiferromagnetic moment or hyperfine coupling, which contributes to the peak-width of $1/T_1^{\rm{CR}}$. Thus our fitting most likely underestimates     
$\tau_{\alpha}$.  

In non-magnetic solids, the nuclear spin-spin coupling is caused by dipolar field, leading to the following expression of the  second moment\cite{Abragam}
\begin{equation}
\braket{\Delta\omega^2}_{\rm{dip}}=\frac{(2\pi)^2}{3}\gamma^4 h^2I(I+1)\sum_k\frac{(1-3\cos^2\theta_k)^2}{r^6_k} . 
\end{equation}
This expression has to be modified in our case since the direction of the local field is not the same for V$^{4+}$ and V$^{5+}$ nuclei. After a straightforward but tedious procedure to reexpress Eq.~(\ref{spin-spin}) using a common coordinate frame for V$^{4+}$ and V$^{5+}$ nuclear spins, the second moment due to the dipolar coupling is calculated to be   
$\sqrt{\braket{\Delta\omega^2}_{\rm dip}} = 2 \times 10^{3}$ s$^{-1}$. Therefore, we conclude that the coupling between V$^{4+}$ and V$^{5+}$ nuclei in the AF state of K$_2$V$_3$O$_8$ is 50 times stronger than the dipolar coupling. 

In magnetic materials, strongly enhanced nuclear spin-spin coupling has been observed and a few mechanisms are known where such coupling is mediated by electronic processes, for example, the Ruderman-Kittel-Kasuya-Yoshida type interaction in $f$-electron metals\cite{Tokunaga, Tokunaga2, Chudo} or virtual spin wave excitations in insulating spin systems.\cite{Suhl, Nakamura} What is particularly remarkable in our case is that the highly enhanced cross relaxation rate $1/T_1^{\rm{CR}}$ at the peak frequency is nearly independent of temperature below $T_{\rm N}$ down to the lowest temperature of our measurements (see $(1/T_1)_c$ at 6.615~T in Fig.~\ref{fig:T1_temp}), in strong contrast to the rapid suppression of $1/T_1$ when unaffected by cross relaxation mechanism. This indicates persistence of gapless spin fluctuations in the ordered phase at high magnetic fields. Gapless spin excitations at high magnetic fields have been also proposed to be responsible for the highly enhanced thermal conductivity.\cite{Sales}  Such gapless fluctuations could be related to phase modes in incommensurate spin structures or complex texture.

\section{Conclusion}
The $^{51}$V NMR measurements on non-magnetic ${\rm{V}}^{5+}$ sites in the quasi 2D antiferromagnet K$_2$V$_3$O$_8$ revealed complex line splitting and subsequent line broadening below the structural transition temperature (115 K) due to significant change of the hyperfine coupling caused by local lattice distortion. Crude feature of the NMR spectra with broad double peak structure for the magnetic field along the $c$-axis in the antiferromagnetically ordered state is qualitatively explained by a simple N\'{e}el order with a spin flop transition, although precise determination of the spin structure is not possible due to structural complexity. In contrast to the rather conventional static behavior, dynamic anomaly revealed by huge enhancement of the nuclear spin-lattice relation rate $1/T_1$ in a certain range of magnetic fields points to cross relaxation caused by extremely strong nuclear spin-spin coupling between non-magnetic ${\rm{V}}^{5+}$ and magnetic ${\rm{V}}^{4+}$ sites. This might be closely associated with the strong gapless spin fluctuations associated with possible incommensurate spin structure or exotic spin texture. 

\begin{acknowledgments}
This study was supported by the JSPS KAKENHI Grant Numbers JP17H02918, JP25287083 and JP18H04310 (J-Physics). Work at ORNL was supported by the US Department of Energy, Office of Science, Basic Energy Sciences, Materials Sciences and Engineering Division.
\end{acknowledgments}

\end{document}